\documentclass[aps,amsfonts,amssymb,pra,preprint]{revtex4} 
\newtheorem{theorem}{Theorem}
\newtheorem{corollary}{Corollary}
\def\qed{$\Box$}
\newtheorem{definition}{Definition}
\newtheorem{proof}{Proof}
\def\proof{{\bf Proof:~}}
\usepackage{graphicx}
\usepackage{amssymb}

\begin{document}

\title{A combinatorial  approach for studying  LOCC transformations of
multipartite       states}      \author{Sudhir       Kumar      Singh}
\affiliation{Department  of   Electrical  Engineering,  University  of
California,   Los    Angeles,   CA   90095}   \email{suds@ee.ucla.edu}
\author{Sudebkumar  Prasant Pal}  \affiliation{Department  of Computer
Science  and Engineering, Indian  Institute of  Technology, Kharagpur-
721302,  India} \email{spp@cse.iitkgp.ernet.in}  \author{Somesh Kumar}
\affiliation{Department   of    Mathematics,   Indian   Institute   of
Technology,            Kharagpur-            721302,            India}
\email{smsh@maths.iitkgp.ernet.in}        \author{R.         Srikanth}
\affiliation{Raman  Research  Institute,  Sadashiva Nagar,  Bangalore-
560080, Karnataka, India} \email{srik@rri.res.in}


\begin{abstract}
We develop  graph theoretic methods for  analysing maximally entangled
pure  states distributed between  a number  of different  parties.  We
introduce  a technique called  {\it bicolored  merging}, based  on the
monotonicity  feature   of  entanglement  measures,   for  determining
combinatorial conditions  that must be satisfied for  any two distinct
multiparticle  states  to be  comparable  under  local operations  and
classical communication  (LOCC).  We present several  results based on
the possibility or impossibility of comparability of pure multipartite
states.   We show  that there  are exponentially  many  such entangled
multipartite states among $n$ agents.  Further, we discuss a new graph
theoretic  metric  on  a   class  of  multi-partite  states,  and  its
implications.
\end{abstract}

\maketitle

\section{Introduction}
Given  the extensive  use of  quantum entanglement  as a  resource for
quantum information  processing \cite{ekertpoi,nc2000}, the  theory of
entanglement, in  particular, entanglement quantification,  is a topic
important  to  quantum  information  theory.  However,  apart  from  a
limited number of cases like low dimension Hilbert spaces and for pure
states, the  mathematical structure of  entanglement is not  yet fully
understood. The entanglement properties  of bipartite states have been
widely explored (see \cite{brub02,hord01} for a comprehensive review).
This has been aided by the fact that bipartite states possess the nice
mathematical  property  in  the  form  of  the  Schmidt  decomposition
\cite{nc2000},  the   Schmidt  coefficients  encompassing   all  their
non-local properties.  No such simplifying  structure is known  in the
case of  larger systems.  Approaches using certain  generalizations of
Schmidt  decomposition  \cite{bennet2000,kempe99,partovi04} and  group
theoretic  or algebraic  methods  \cite{linden98,linden99,liu04}, have
been taken  in this  direction. A number  of methods for  comparing or
quantifying  or   qualifying  entanglement  have   been  proposed  for
bipartite systems and/or pure states such as entanglement of formation
\cite{entfor},  entanglement  cost \cite{entfor,entcost},  distillable
entanglement \cite{entfor,entdist2},  relative entropy of entanglement
\cite{relent}, negativity \cite{negent}, concurrence \cite{conent} and
entanglement witnesses  \cite{entwit}.  However, these quantifications
do  not always  lend  themselves  to being  computed,  except in  some
restricted situations. As such, a general formulation is still an open
problem.

It  is known  that state  transformations under  local  operations and
classical  communication  (LOCC)  are  very important  to  quantifying
entanglement  because LOCC  can at  the best  increase  only classical
correlations. Therefore a good measure of entanglement is expected not
to increase under  LOCC. A necessary and sufficient  condition for the
possibility of  such transformations in  the case of  bipartite states
was given by Nielsen \cite{neilsen99}. An immediate consequence of his
result was the existence of {\it incomparable} states (the states that
can  not   be  obtained  by   LOCC  from  one  another).   Bennett  et
al.  \cite{bennet2000},   formalized  the  notions   of  reducibility,
equivalence  and incomparability  to multi-partite  states and  gave a
sufficient  condition  for  incomparability  based  on  {\it  partial}
entropic criteria.

In this work,  our principal aim is not  to quantify entanglement, but
to develop graph theoretic  techniques to analyze the comparability of
maximally entangled multipartite  states of several qubits distributed
between a number of  different parties.  We obtain various qualitative
results  concerning reversibility of  operations and  comparability of
states by  observing the combinatorics  of multiparitite entanglement.
For  our purpose,  it is  sufficient to  consider the  graph theoretic
representation of  various maximally entangled  states (represented by
specific graphs built  from EPR, GHZ and so  on).  Although this might
at  first  seem  overly  restrictive,  we  will in  fact  be  able  to
demonstrate a number of new results.  Furthermore, being based only on
the  monotonicity  principle,  it  can  be  adapted  to  any  specific
quantification  of  entanglement.  Therefore,  our approach  is  quite
generic, in principle applicable  to all entanglement measures.  Since
the entanglement of maximally  entangled states is usually represented
by integer values, it turns  out that we can analyze entangled systems
simply  by studying  the combinatorial  properties of  graphs  and set
systems representing  the states.  The basic definitions  and concepts
are introduced  through the framework set  in Section \ref{framework}.
We  introduce a technique  called {\it  bicolored merging}  in Section
\ref{bicoloredmerging},  which is essentially  a combinatorial  way of
quantifying maximal  entanglment between two parts of  the system, and
inferring transformation properties to be satisfied by the states.

In   Section  \ref{skpsel},   we   present  our   first  result:   the
impossibility  of obtaining  two  Einstein-Podolsky-Rosen (EPR)  pairs
among three players  starting from a Greenberger-Horne-Zeilinger (GHZ)
state  (Theorem \ref{skpghzreverse}). We  then show  that this  can be
used  to establish  the  impossibility of  implementing a  two-pronged
teleportation (called {\it  selective teleportation}) given pre-shared
entanglement in the form of  a GHZ state.  We then demonstrate various
classes    of   incomparable    multi-partite   states    in   Section
\ref{classmulentan}. Finally, we discuss  the minimum number of copies
of  a state  required to  prepare another  state by  LOCC  and present
bounds on this  number in terms of the  {\it quantum distance} between
the two states in Section \ref{quantumdistance}.

We believe that our combinatorial approach vastly simplifies the study
of entanglement  in very  complex systems. Moreover,  it opens  up the
road  for further  analysis,  for example,  to interpret  entanglement
topologically.  In future  works, we intend to apply  and extend these
insights to non-maximal and  mixed multipartite states, and to combine
our approach with a suitable measure of entanglement.

\section{The Combinatorial Framework}
\label{framework}

In this section we introduce a number of basic concepts useful to describe
combinatorics of entanglement. 
First, an {\em EPR graph} $G(V, E)$ is a graph whose vertices are the players
($\in V$) and edges ($\in E$)
represent shared entanglement in the form of an EPR pair. Formally:
\begin{definition}
EPR graph: For $n$ agents $A_1, A_2, \cdots , A_n$ an undirected graph $G = (V,E)$
is constructed as
follows:
$ V = \{ A_i: i=1, 2, \cdots , n \}$ ,
$ E = \{ \{ A_i, A_j \}: A_i $ and  $A_j ~\mbox{share an EPR pair}, 1 \leq i, j \leq n; i \neq j \}$.
The graph $G = (V, E)$ thus formed is called the EPR graph of the $n$ agents.
\end{definition} 

A spanning tree is a graph which connects all vertices without forming cycles (i.e., loops).
Accordingly:
\begin{definition} Spanning EPR tree: 
A {\em spanning tree} is a connected, undirected graph linking all vertices 
without forming cycles. 
An EPR graph $G = (V, E)$ is called a {\em spanning EPR tree} if the undirected graph 
$G = (V, E) $ is a spanning tree. 
\end{definition}

The  above  notions  are  generalized  to  more  general  multipartite
entanglement by  means of the concept  of a {\em  hypergraph}. A usual
graph is built up from edges,  where a normal edge links precisely two
vertices.  A hyperedge is  a generalization  that links  $r$ vertices,
where $r \ge 2$. A graph endowed with at least one hyperedge is called
a  hypergraph.   From  the   combinatorial  viewpoint,  a  simple  and
interesting   connection  can   be  made   between   entanglement  and
hyperedges: an $n$-cat state  (also sometimes called an $n$-GHZ state)
corresponds to  a hyperedge of size  $n$. In particular,  an EPR state
corresponds to a simple edge connecting only two vertices. Formally:
\begin{definition}
Entangled hypergraph: Let $S$ be the  set of $n$ agents and $F= \{E_1,
E_2, \cdots , E_m \}$, where $ E_i  \subseteq S; i = 1, 2, \cdots , m$
and $E_i$ is such that its elements (agents) are in $|E_i|$-CAT state.
The hypergraph  (set system)  $ H =  (S, F)  $ is called  an entangled
hypergraph of the $n$ agents.
\end{definition} 
A graph  is connected if there  is a path  (having a length of  one or
more edges) between any two vertices. Accordingly:
\begin{definition}
Connected entangled  hypergraph: A  sequence of $j$  hyperedges $E_1$,
$E_2$,  ...,  $E_j$  in  a  hypergraph  $H=(S,F)$  is  called  a  {\it
hyperpath} (path) from a vertex $a$ to a vertex $b$ if
\begin{enumerate}
\item $E_i$  and $E_{i+1}$ have a  common vertex for  all $1\leq i\leq
j-1$,
\item $a$ and $b$ are agents in $S$,
\item $a\in E_1$, and
\item $b\in E_j$.
\end{enumerate}
If there is  a hyperpath between every pair of vertices  of $S$ in the
hypergraph $H$, we say that $H$ is connected.
\end{definition} 

Analogous to a spanning EPR tree we have:
\begin{definition}
Entangled hypertree: A connected entangled  hypergraph $H = (S, F)$ is
called an entangled hypertree if it contains no cycles, that is, there
do not  exist any pair  of vertices from  $S$ such that there  are two
distinct paths between them.
\end{definition}
Further:
\begin{definition}
$r$-uniform entangled hypertree: 
An entangled hypertree is called an $r$-uniform entangled hypertree 
if all of its hyperedges are of size $r$ for $r\geq 2$.
\end{definition}

In ordinary  graphs, a vertex  that terminates, i.e., has  precisely a
single edge linked to it is  called a terminal or pendent vertex. This
concept is extended to the case of hypergraphs:
\begin{definition}
Pendant  Vertex: A  vertex  of  a hypergraph  $H=(S,F)$  such that  it
belongs to  only one hyperedge  of $F$ is  called a pendant  vertex in
$H$.   Vertices which belong  to more  than one  hyperedge of  $H$ are
called non-pendant.
\end{definition}

In the paper we use polygons for pictorially representing an entangled
hypergraph of multipartite states.  (There should be no confusion with
a closed  loop of EPR pairs  because we consider  only tree structured
states).   A hyperedge  representing  an $n$-CAT  amongst the  parties
$\{i_1, i_2,  \cdots, i_n\}$ is pictorially represented  by an $n$-gon
with  vertices distinctly  numbered by  $i_1, i_2,  \cdots,  i_n$.  We
write these vertices $i_1, i_2,  \cdots, i_n$ corresponding to the $n$
vertices of  the $n$-gon in the pictorial  representation in arbitrary
order. This  only means  that out  of $n$ qubits  of the  $n$-CAT, one
qubit is with each of the $n$ parties.

A result we will require  frequently is that there exist teleportation
\cite{bennett} protocols to  produce $n$-partite entanglement starting
from pairwise  entanglement shared along any  spanning tree connecting
the  $n$  parties. That  is,  there exist  LOCC  protocols  to turn  a
$n$-party spanning EPR tree  into an $n$-regular hypergraph consisting
of  a single  hyperedge  of size  $n$.   The protocol  is detailed  in
Ref.  \cite{suds}, but  the basic  idea  is readily  described. It  is
essentially a scheme to deterministically create a maximally entangled
$n$-cat  state   from  $n-1$  EPR   pairs  shared  along   a  spanning
tree. Briefly, the protocol consists in teleporting entanglement along
a  spanning tree.  Players not  on  terminal vertices  along the  tree
execute  the  following subroutine.  Suppose  player  Alice shares  an
$m$-cat with  $(m-1)$ preceding players  along the tree and  wishes to
create an  $(m+1)$-cat state including  Bob, the next player  down the
tree. First she  entangles an auxiliary particle with  her particle in
the $m$-cat state  by means of local operation. She  then uses her EPR
pair shared with  Bob to teleport the state  of the auxiliary particle
to Bob.   The $(m+1)$ players, including  Alice and Bob,  now share an
$(m+1)$-cat state, as desired.

Another result  we will require  in some of  our proofs, given  as the
theorem below, is that the spanning EPR tree mentioned above is also a
necessary condition  to prepare an $n$-CAT state  starting from shared
EPR pairs.
\begin{theorem} 
\label{skpnec}
 Given  a communication  network of  $n$  agents with  only EPR  pairs
 permitted  for  pairwise  entanglement  between agents,  a  necessary
 condition for  creation of a $n$-CAT  state is that the  EPR graph of
 the $n$ agents must be connected.
\end{theorem}
Proof  of the  theorem is  given  in Appendix  1 using  our method  of
bicolored merging developed in section III.

\section{Bicolored Merging}
\label{bicoloredmerging}

Monotonicity is  easily the most natural characteristic  that ought to
be satisfied  by all entanglement measures  \cite{hord01}. It requires
that  any appropriate measure  of entanglement  must not  change under
local unitary operations and more generally, the expected entanglement
must not  increase under LOCC.  We should note  here that in  LOCC, LO
involves  unitary  transformations, additions  of  ancillas (that  is,
enlarging the Hilbert Space), measurements, and throwing away parts of
the system, each of these actions performed by one party on his or her
subsystem.  CC between  the parties allows local actions  by one party
to  be  conditioned  on  the  outcomes  of  the  earlier  measurements
performed by the other parties.

\begin{figure}
\begin{center}
\scalebox{0.5}{\includegraphics{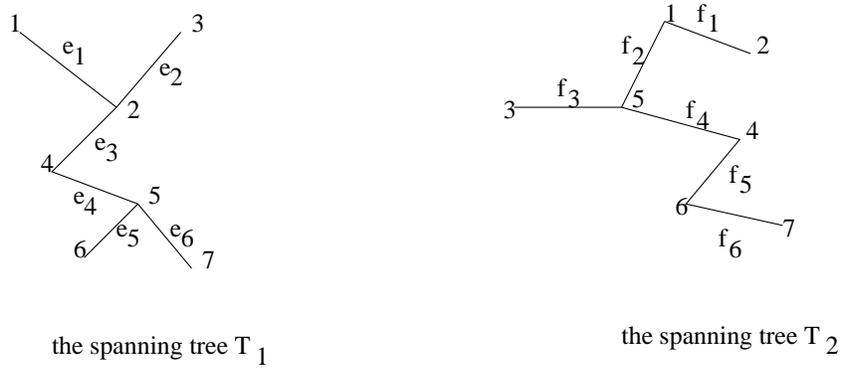}}
\end{center}
\caption{Spanning EPR trees $T_1$ and $T_2$}
\label{figure30A}
\end{figure}

Apart  from  monotonicity,  there  are certain  other  characteristics
required   to  be   satisfied  by   entanglement   measures.  However,
monotonicity  itself  vastly  restricts  the  choice  of  entanglement
measures (for  example, marginal entropy as a  measure of entanglement
for  bipartite pure  states  or entanglement  of  formation for  mixed
states).  In the present work, we find that monotonicity, where proven
for  a particular  entanglement measure  candidate, restricts  a large
number of state  transformations and gives rise to  several classes of
incomparable  (multi-partite)  states.   So,  in order  to  study  the
possible state  transformations of (multi-partite)  states under LOCC,
it would be interesting to look  at the kind of state transforms under
LOCC  which  monotonicity  does   not  allow.   We  can  observe  that
monotonicity does not allow the preparation of $n+1$ or more EPR pairs
between two parties starting from only $n$ EPR pairs between them.  In
particular,  it is  not  possible to  prepare  two or  more EPR  pairs
between  two parties starting  only with  a single  EPR pair  and only
LOCC.   This  is an  example  of  impossible  state transformation  in
bipartite  case  as  dictated   by  the  monotonicity  postulate.   We
anticipate that  a large class  of multi-partite states could  also be
shown  to  be incomparable  by  using  impossibility  results for  the
bipartite  case through suitable  reductions.  For  instance, consider
transforming  (under LOCC)  the state  represented by  a  spanning EPR
tree, say $T_1$, to that  of the state represented by another spanning
EPR tree, say $T_2$ (See Figure \ref{figure30A}).  This transformation
can be  shown to be  impossible by reducing  to the bipartite  case as
follows: We assume  for the sake of contradiction  that there exists a
protocol  $P$ which can  perform the  required transformation.   It is
easy to see that the protocol  $P$ is also applicable in the case when
a party $A$  possesses all the qubits  of parties $4, 5, 6,  $ and $7$
and another party $B$ possesses all the qubits of the parties $1, 2, $
and $3$.  This means that party $A$ is playing the role of parties $4,
5, 6, $  and $7$ and $B$ is  playing the role of parties $1,  2, $ and
$3$.   Clearly,  any  LOCC   actions  done  within  group  $\{1,2,3\}$
($\{4,5,6,7\}$) is  a subset of LO  available to $B$ ($A$)  and any CC
done  between   one  party  from   $\{1,2,3\}$  and  the   other  from
$\{4,5,6,7\}$ is managed by CC between $B$ and $A$.

\begin{figure}
\begin{center}
\scalebox{0.2}{\includegraphics{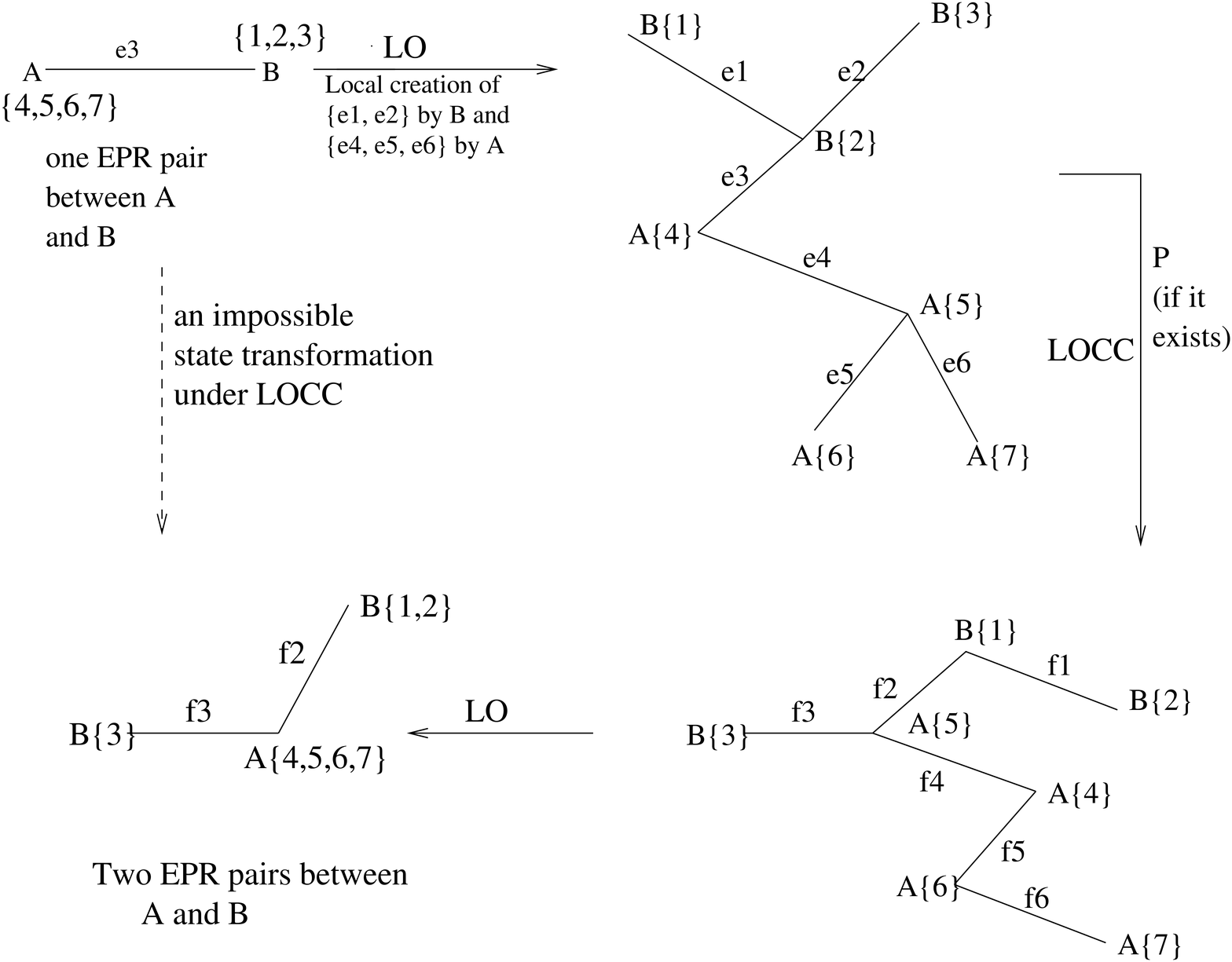}}
\end{center}
\caption{Converting $T_1$ to $T_2$ under LOCC}
\label{figure30C}
\end{figure}

Therefore,  starting  only  with  one  edge  ($e_3$)  they  eventually
construct  $T_1$  just   by  LO  (by  local  creation   of  EPR  pairs
representing  the edges  $e_1, e_2,  e_4,  e_5, $  and $e_6$  ($\{e_1,
e_2\}$  by  $B$ and  $\{e_4,  e_5, e_6\}$  by  $A$).  They then  apply
protocol $P$ to obtain $T_2$ with  the edges $f_1, f_2, f_3, f_4, f_5$
and $f_6$.  (Refer to the  Figure \ref{figure30C}).  All  edges except
$f_2$ and $f_3$ are local EPR pairs (that is, both qubits are with the
same party,  $A$ or $B$).  Now the  parties $A$ and $B$  share two EPR
pairs  in the  form of  the edges  $f_2$ and  $f_3$, even  though they
started sharing only one EPR  pair.  But this is in contradiction with
monotonicity:  that expected  entanglement should  not  increase under
LOCC.  Hence, we can conclude that such a protocol $P$ cannot exist!

The approach we took in the above example could also be motivated from
the marginal entropic criterion (noting that this criterion in essence
is  also a  direct implication  of monotonicity).   As clear  from the
above example, the above scheme aims to create a bipartition among the
$n$ players in such a way  that the marginal entropy of each partition
is different for  the two states. In many  cases, this difference will
simply correspond to different number  of EPR pairs shared between the
two partitions. Given two  multi-partite states, the relevant question
is: ``is  there a bipartition such  that the marginal  entropy for the
two states  is different?". If  yes, then the state  (configuration of
entanglement) corresponding  to the higher entropy  cannot be obtained
from that  to lower  entropy by  means of LOCC.   It is  convenient to
imagine the two partitions  being `colored' distinctly to identify the
partitions which they make up.

In general, suppose we 
want to show that the  multi-partite state $|\psi\rangle$ 
can not be converted to the multi-partite state $|\phi\rangle$ by LOCC.
This can be done by showing an assignment of the qubits (of all parties) 
to only two parties such that $|\psi\rangle$ can be obtained from $n$ ($n = 0, 1, 2, \cdots $)
EPR pairs between the two parties by LOCC while $|\phi\rangle$ can be converted 
to more than $n$ EPR pairs between the two parties by LOCC.
This is equivalent to saying that each party is given either of two colors (say $A$ or $B$).
Finally all qubits with parties colored with color $A$ are assigned to
the first party (say $A$) and that with parties colored with second color to
the second party (say $B$). This coloring is done in such a way that the state $|\psi\rangle$ 
can be obtained by LOCC from less number of EPR pairs between $A$ and $B$ than that can be obtained from
$|\phi\rangle$ by LOCC.
Local preparation (or throwing away) of EPR pairs is what we call merging in combinatorial sense.  
Keeping this idea in mind, we now formally introduce the idea of 
bicolored merging for such reductions in the case of 
the multi-partite states represented by EPR graphs and entangled hypergraphs. 

Suppose that there are two EPR graphs $G_1=(V, E_1)$ and $G_2=(V, E_2)$ on the same 
vertex set $V$ (this means that these two multi-partite states 
are shared amongst the same set of parties) and we want to 
show the impossibility of transforming 
$G_1$ to $G_2$ under LOCC, then this is reduced 
to a bipartite LOCC transformation 
which violates monotonicity, as follows:

\begin{enumerate}
\item Bicoloring: Assign either of the two colors $A$ or $B$ to every vertex, that is, 
each element of $V$. 
\item Merging: For each element $\{v_i, v_j\}$ of $E_1$, merge the two vertices $v_i$ and 
$v_j$ if and only if they have been assigned the same color during the bicoloring stage and assign 
the same color to the merged vertex. Call this graph obtained from $G_1$ as BCM (Bicolored-Merged) EPR graph of 
$G_1$ and denote it by $G_1^{bcm}$. Similarily, obtain the BCM EPR graph $G_2^{bcm}$ of 
$G_2$.
\item The bicoloring and merging is done in such a way that the graph $G_2^{bcm}$ has 
more number of edges than that of $G_1^{bcm}$. 
\item Give all the qubits possessed by the vertices with color $A$ to the first party (say, party $A$)
and all the qubits possessed by the vertices with color $B$ to the second party (say, party $B$).
Combining this with the previous steps, it is ensured that in the bipartite reduction of the multi-partite state
represented by $G_2$, the two parties $A$ and $B$ share more number of EPR pairs (say, state $|\psi_2\rangle$)
than that for 
$G_1$ (say, state $|\psi_1\rangle$).  
\end{enumerate}

\begin{figure}
\scalebox{0.3}{\includegraphics{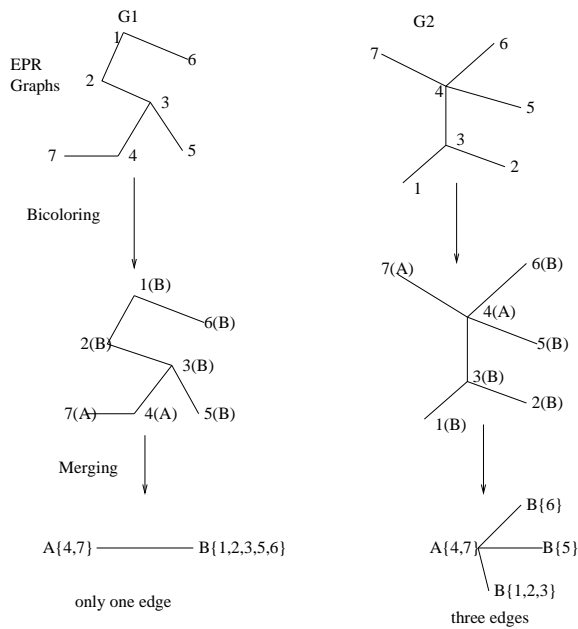}}
\caption{Bicolored merging of EPR graphs}
\label{figure31}
\end{figure}

We denote this reduction as $G_1\ngtr G_2$.
Now if there exits a protocol $P$ which can transform $G_1$ to $G_2$ by LOCC, then 
$P$ can also transform $|\psi_1\rangle$ to $|\psi_2\rangle$ just by LOCC as follows:
$A$ ($B$) will play the role of all vertices in $V$ which were colored as $A$ ($B$).
The edges which were removed due to merging can easily be cretated by local operations 
(local preparation of EPR pairs) by the party $A$ ($B$) if the color of the merged end-vertices 
of the edge was assigned color $A$ ($B$). This means that starting from $|\psi_1\rangle$
and only LO, $G_1$ can be created. This graph is virtually amongst $|V|$ parties even though 
there are only two parties. The protocol $P$ then, can be applied to $G_1$ to obtain $G_2$ by LOCC.
Subsequently $|\psi_2\rangle$ can be obtained by the necessary merging of vertices by LO, that is by 
throwing away the local EPR pair represented by the edges between the vertices being merged.
Since the preparation of $|\psi_2\rangle$ from $|\psi_1\rangle$ by LOCC violates the monotonocity postulate,
such a protocol $P$ can not exist! An example of bicolored merging for EPR graphs has been 
illustrated in Figure \ref{figure31}.

\begin{figure}
\scalebox{0.4}{\includegraphics{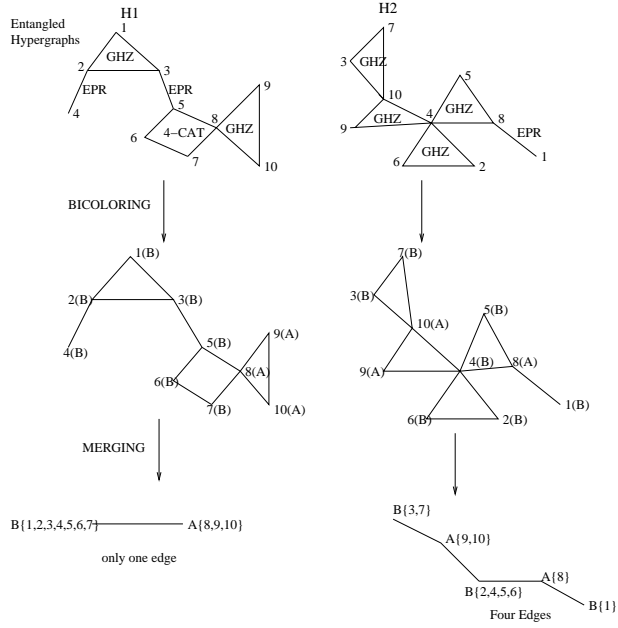}}
\caption{Bicolored merging of entangled hypergraphs}
\label{figure32}
\end{figure}

The bicolored merging in the case of entangled hypergraphs is essentially the same as that for 
EPR graphs. For the sake of completeness, we present it here.
Suppose there are two entangled hypergraphs $H_1=(S, F_1)$ and $H_2=(S, F_2)$ on the same 
vertex set $S$ (that is, the two multi-partite states are shared amongst the 
same set of parties) and we want to show the impossibility of transforming 
$H_1$ to $H_2$ under LOCC. Transformation of $H_1$ to $H_2$ can be reduced to a bipartite LOCC transformation 
which violates monotonicity thus proving the impossibility. The reduction is done as follows:

\begin{enumerate}
\item Bicoloring: Assign either of the two colors $A$ or $B$ to every vertex, 
that is, each element of $S$. 
\item Merging: For each element $E=\{v_{i1}, v_{i2}, \cdots ,  v_{ij}\}$ of $F_1(F_2)$, merge 
all vertices with color $A$ to one vertex and those with color $B$ to another vertex and give 
them colors $A$ and $B$ respectively. This merging collapses each hyperedge to either a simple 
edge or a vertex and thus the hypergraph reduces to a simple graph with vertices assigned with either 
of the two colors $A$ or $B$.  
Call this graph obtained from $H_1$ as BCM EPR graph of 
$H_1$ and denote it by $H_1^{bcm}$. Similarily obtain the BCM EPR graph $H_2^{bcm}$ of 
$H_2$.
\item The bicoloring and merging is done in such a way that the graph $H_2^{bcm}$ has 
more number of edges than that of $H_1^{bcm}$. 
\item Give all the qubits possessed by the vertices with color $A$ to the party one (say party $A$)
and all the qubits possessed by the vertices with color $B$ to the second party (say party $B$).
\end{enumerate}

We denote the above reduction as $H_1\ngtr H_2$.
The rest of the discussion is similar to that for the case
of EPR graphs given before. In the Figure \ref{figure32},
we demostrate the bicolored merging of entangled hypergraphs. 
Note that the two entangled 
hypergraphs $H_1$ and $H_2$ are LOCC comparable only if one of 
$H_1\ngtr H_2$ and $H_2\ngtr H_1$ 
is not true. Equivalently, if both of $H_1\ngtr H_2$ and
$H_2\ngtr H_1$ hold, then  the entangled  
hypergraphs $H_1$ and $H_2$ are incomparable. 

It is also 
interesting to note at this point that LOCC incomparability shown
by using the method of bicolored merging is in fact {\it 
strong incomparability} as defined in \cite{vwani02}.
We would also like to stress that any kind of reduction (in particular, 
various possible extensions of bicolored merging) which leads to the violation of {\it any} of
the properties of a potential entanglement measure, is pertinent to show the impossibility
of many multi-partite state transformations under LOCC. Since the bipartite case has been extensively studied,
such reductions can potentially provide many ideas about multi-partite case by just exploiting 
the results from the bipartite case. In particular, the definitions of EPR graphs and entangled hypergraphs
could also be suitably extended to capture more types of multi-partite pure states and even mixed states 
and a generalization of the idea of bicolored merging as a suitable reduction for this case could also
be worked out. 

\section{LOCC Incomparability and Selective Teleportation}
\label{skpsel}

We know that a GHZ state amongst three agents $A$, $B$ and $C$ 
can be prepared from EPR pairs shared between any two pairs of
the three agents using only LOCC \cite{suds,bos98,zei97,zuk95}. 
We consider the problem of {\it reversing} 
this operation, that is, whether it is possible
to construct two EPR pairs between 
any two pairs of the three agents from a GHZ state amongst the three 
agents, using only LOCC. By using the 
method of bicolored merging, we answer this question
in the negative by
establishing the following theorem.

\begin{theorem}
\label{skpghzreverse}
Starting from a GHZ state shared amongst three parties in a 
communication network, two EPR pairs cannot be created 
between any two sets of two parties using only LOCC.
\end{theorem}

\begin{figure}
\scalebox{0.5}{\includegraphics{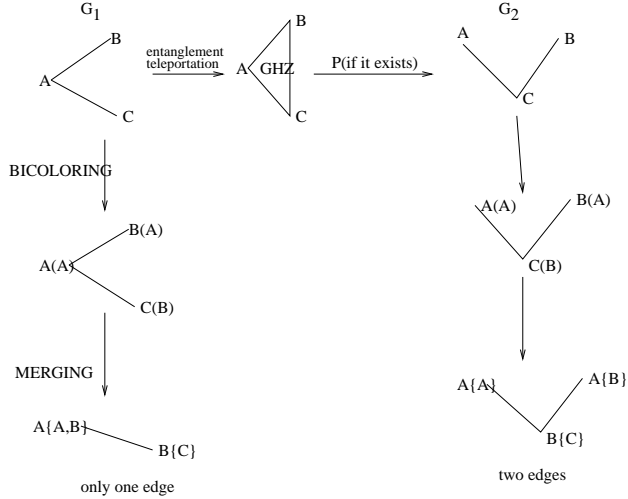}}
\caption{LOCC-irreversibility of the process [2 EPR $\rightarrow$ GHZ].}
\label{figure33}
\end{figure}

\proof
Suppose there exists a protocol $P$ for {\it reversing} a GHZ state  
into two EPR pairs using only LOCC. In particular, suppose
protocol $P$ starts with a GHZ state amongst the agents 
$A$, $B$ and $C$, and prepares EPR pairs between any two pairs of  
$A$, $B$ and $C$ (say, $\{A,C\}$, and $\{B,C\}$, corresponding to configuration $G_1$ as shown in
Figure \ref{figure33}).
Since we can prepare the GHZ state from EPR pairs between any two pairs
of the three agents, we can prepare the GHZ state starting from EPR pairs
between $A$ and $B$, and $A$ and $C$. Once the GHZ state is prepared, we 
can apply protocol $P$ to construct EPR pairs between $A$ and $C$ and between $B$
and $C$ using only LOCC (i.e., configuration $G_2 \equiv \{\{A,C\},\{B,C\}\}$).
So, we can use only LOCC to convert a 
configuration where EPR pairs exist between $A$ and $C$ and between $A$ and $B$, to
a configuration where EPR pairs are shared between $A$ and $C$ and between $B$ and $C$.
The possibility of $P$ means that
the marginal entropy of $C$ can be increased using only LOCC, which is known to be
impossible. \hfill \qed


The same result could also be achieved by similar bicolored merging
directly applied on the GHZ state and any of $G_1$ or $G_2$ but
we prefer the above proof for stressing the argument on the symmetry 
of $G_1$ and $G_2$ with respect to the GHZ. Moreover,
this proof gives an intuition about possibility of incomparability
amongst spanning EPR trees as $G_1$ and $G_2$ are two
distinct spanning EPR trees on three vertices.
We prove this general result in the Theorem \ref{twoeprtrees}.   

The above theorem motivates us to propose some kind of comparison between 
a GHZ state and two pairs of EPR pairs in terms of the 
non-local correlations they possess. In this sense, therefore, 
a GHZ state may be viewed as less than two EPR pairs.
It is easy to see that an EPR pair between
any two parties can be obtained starting only 
from a GHZ state shared amongst the three parties and 
LOCC. The third party will just do a measurement in
the diagonal basis and send
the result to other two. By applying the corresponding 
suitable operations they get the required EPR pair.
>From Theorem \ref{skpnec}, we observe that a single EPR pair, between any
two of the three parties, is not sufficient for preparing a GHZ state
amongst the three parties using only LOCC.
These arguments can be summarised in the following 
theorem.

\begin{theorem}
\label{eprghz}

1-EPR pair $<_{LOCC}$ a GHZ state $<_{LOCC}$ 2-EPR pairs 

\end{theorem}

An interesting problem in quantum information theory is that of
{\it selective teleportation} \cite{mks}. Given three agents $A$, $B$ 
and $C$, and two qubits of unknown quantum states 
$|\psi_1\rangle$ 
and
$|\psi_2\rangle$ with $A$, the problem is to send 
$|\psi_1\rangle$ to $B$ and  
$|\psi_2\rangle$ to $C$ selectively, using only LOCC and
apriori entanglement between the three agents.
A simple solution to this problem is applying standard 
teleportation \cite{bennett}, in the case where $A$ shares 
EPR pairs with both $B$ and $C$.
An interesting question is whether any other 
form of apriori entanglement
can help achieving selective teleportation. In particular,
is it possible to perform selective teleportation where 
the apriori entanglement is in the form of a GHZ state amongst the 
three agents? The following theorem answers this question using the 
result of the Theorem \ref{skpghzreverse}.

\begin{theorem}
\label{seltel}
With apriori entanglement given in the form of a
GHZ state shared amongst
three agents, two qubits can not be
selectively teleported by ane of the three
parties to the other two parties. 
\end{theorem}

\begin{proof}
Suppose there exists a protocol $P$ which
can enable one of the three parties (say $A$)
to teleport two qubits $|\psi_1\rangle$ and
$|\psi_2\rangle$ selectively to the other two parties
(say $B$ and $C$).
Now $A$ takes four qubits; she prepares two EPR pairs
one from the first and second qubits and the other from
the third and fourth qubits. He then teleports the first
and third qubits selectively to $B$ and $C$ using $P$
( consider first qubit as $|\psi_1\rangle$ and the third
qubit as $|\psi_2\rangle$ ). We can note here that
in this way $A$ is able to share one EPR pair each
with $B$ and $C$. But this is impossible because it allows
$A$ to prepare two EPR pairs starting from a GHZ state
and only using LOCC. This contradicts Theorem \ref{skpghzreverse}.
Hence follows the result.
\end{proof}
\hfill \qed

\section{Combinatorial conditions for LOCC incomparability of EPR graphs}  
\label{classmulentan}

An immediate result comparing an $n$-CAT state with EPR pairs follows from 
noting that, given a spanning EPR tree among $n$ parties, an $n$-CAT state
can be constructed using only LOCC using the teleportation protocol described
in Section \ref{framework}. The result we present below generalizes Theorem \ref{eprghz}.

\begin{theorem}
\label{catepr}

$ 1$-EPR  $pair <_{LOCC} n$-CAT $<_{LOCC} (n-1)$-spanning EPR tree.

\end{theorem}

We  can  argue in  a  similar manner  that  an  $n$-CAT state  amongst
$n$-parties can  not be converted  by just using  LOCC to any  form of
entanglement structure  which possesses EPR  pairs between any  two or
more different sets  of two parties.  Assume this  is possible for the
sake of contradiction.   Then the two edges could be  in either of the
two forms: (1) $\{i_1,i_2\}$  and $\{j_1, j_2\}$ and (2) $\{i_1,i_2\}$
and $\{i_2, j_2\}$,  where $i_1, i_2, j_1, j_2$  are all distinct.  In
bicolored-merging  assign the colors  as follows.   In case  (1), give
color $A$ to $i_2$ and $j_2$ and give the color $B$ to the rest of the
vertices.  In case  (2), give color $A$ to $i_2$ and  color $B$ to the
rest  of the  vertices.   Since both  the  cases are  contrary to  our
assumption,   the   assertion   follows.    Moreover,   from   Theorem
\ref{skpnec}  (see Appendix 1  for proof),  no disconnected  EPR graph
would be able  to yield $n$-CAT just by  LOCC.  These two observations
combined together  lead to the  following theorem which  signifies the
fact that these two multi-partite states can not be compared.
  
\begin{theorem}
\label{catgraph}
A CAT state amongst $n$ agents in a communication network
is LOCC incomparable to any disconnected EPR graph associated 
with the $n$ agents having more than one edge.
\end{theorem}

The  above result  indicates that  there  are many  possible forms  of
entanglement  structures  (multi-partite  states)  which  can  not  be
compared at all in terms of non-local correlations they may have. This
simple result  is just an  implication of the  necessary combinatorics
required  for the  preparation of  CAT states.   One  more interesting
question with respect  to this combinatorics is to  compare a spanning
EPR  tree and a  CAT state.   A spanning  EPR tree  is combinatorially
sufficient for preparing  the CAT state and thus  seems to entail more
non-local correlations than in a  CAT state. The question whether this
ordering is  strict needs to be  further investigated.  It  is easy to
see that an EPR pair between  any two parites can be obtained starting
from a CAT  state shared amongst the $n$ agents  just by LOCC (Theorem
\ref{catepr}).  Therefore, given $n-1$ copies  of the CAT state we can
build all the $n-1$ edges of  any spanning EPR tree just by LOCC.  But
whether this  is the lower  bound on the  number of copies  of $n$-CAT
required to obtain an spanning EPR tree is even more interesting.  The
following theorem shows that this is indeed the lower bound.

\begin{theorem}
\label{treecat}  
Starting with only $n-2$ copies of $n$-CAT state shared amongst its 
$n$ agents, no spanning EPR tree of the $n$ agents can be 
obtained just by LOCC. 
\end{theorem}

\begin{figure}
\scalebox{0.5}{\includegraphics{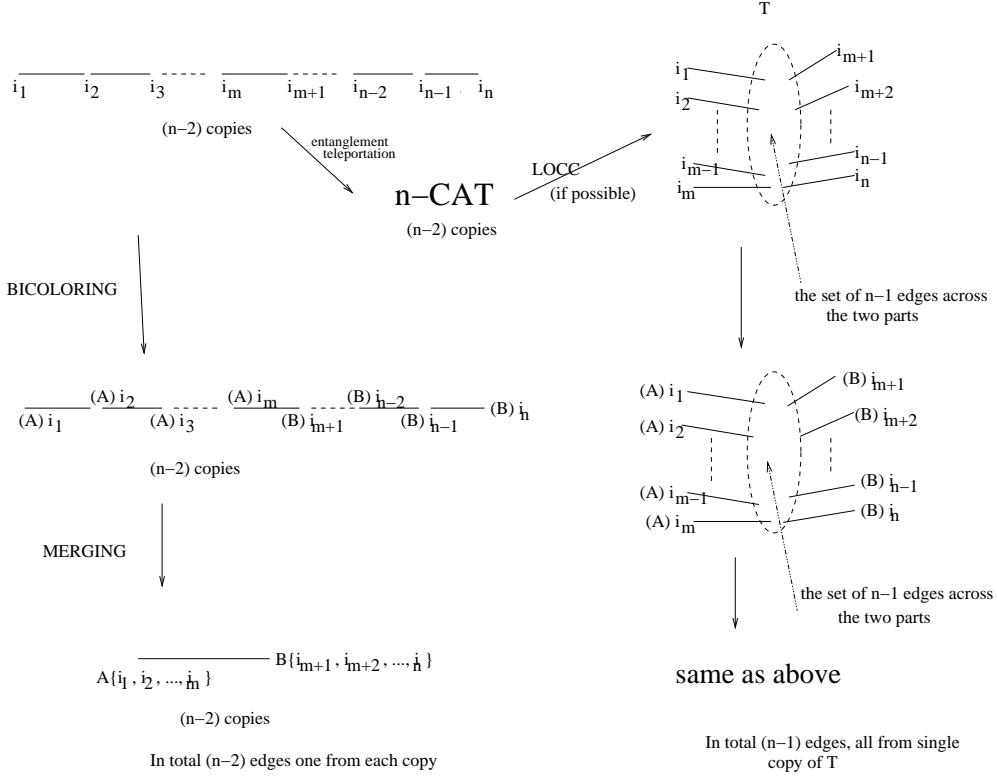}}
\caption{$n-2$ copies of $n$-CAT are not sufficient to prepare an spanning EPR tree }
\label{figure36}
\end{figure}

\begin{proof}
Suppose it is possible to create  a spanning EPR tree $T$ from $(n-2)$
copies  of  $n$-CAT states.   As  we know,  an  $n$-CAT  state can  be
prepared     from     any     spanning     EPR    tree     by     LOCC
\cite{bos98,zei97,zuk95,suds}.  Thus, if $(n-2)$ copies of $n$-CAT can
be converted to $T$, then $(n-2)$  copies of any spanning EPR tree can
be converted to $T$ just by  LOCC.  In particular, $(n-2)$ copies of a
chain  EPR  graph  (which  is  clearly a  spanning  EPR  tree,  Figure
\ref{figure36}) can  be converted to $T$  just by LOCC.   Now, we know
that any tree  is a bipartite connected graph  with $n-1$ edges across
the two parts. Let vertices $i_1,i_2,  \cdots , i_m$ be the members of
the first group and the rest be in the other group.  Construct a chain
EPR graph where the first $m$ vertices are $i_1, i_2, \cdots , i_m$ in
the sequence, and the rest of the vertices are from the other group in
the sequence  (Figure \ref{figure36}).  In bicolored  merging, we give
the color $A$ to the parties  $\{i_1, i_2, \cdots ,i_m\}$ and the rest
of  the parties are  given the  color $B$.   This way  we are  able to
create  $(n-1)$ EPR  pairs (note  that there  are $n-1$  edges  in $T$
across the two groups) between  $A$ and $B$ starting from only $(n-2)$
EPR pairs (considering the  $n-2$ chain-like spanning EPR trees).  So,
we conclude that $(n-2)$ copies of $n$-CAT can not be converted to any
spanning  EPR  tree  just  by  LOCC.  See  Figure  \ref{figure36}  for
illustration of the required  bicolored merging.  The proof could also
be acheived by  similar kind of bicolored merging  directly applied on
$n$-CAT and $T$.
\end{proof} \hfill \qed

In the preceding results we have compared spanning EPR trees with 
CAT states. We discuss the comparability/incomparability of
two distinct spanning EPR trees in the next theorem and corollary .   
 
\begin{theorem}
\label{twoeprtrees}
 Any two distinct spanning  EPR trees are LOCC-incomparable. 
\end{theorem}

\begin{figure}
\scalebox{0.4}{\includegraphics{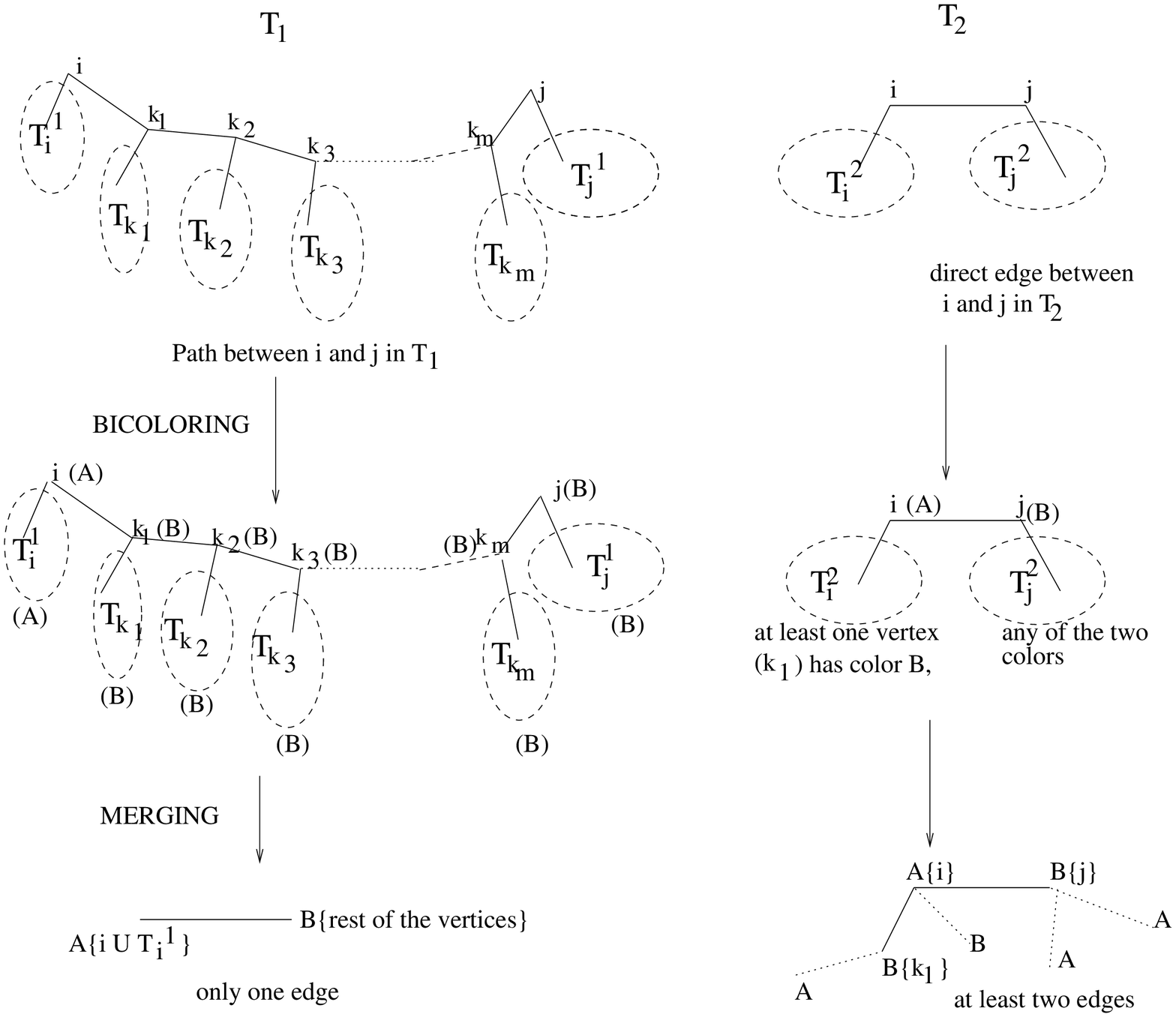}}
\caption{ Spanning EPR trees are LOCC incomparable}
\label{figure37}
\end{figure}

{\it Proof:}
Let $T_1$ and $T_2$ be the two distinct spanning
EPR trees on same $n$ vertices.
Clearly,
there exist two vertices (say $i$ and $j$) which are
connected by an edge in $T_2$ but not in $T_1$.
Also by virtue of connectedness of spanning trees,
there will be a path between $i$ and $j$ in $T_1$.
Let this path be $i k_1 k_2 \cdots k_m j$ with
$m > 0$ (See Figure \ref{figure37}).
Since $m > 0$, $k_1$ must exist.

Let $T_i^{1} \equiv$ subtree in $T_1$ rooted at $i$ except for the branch which
contains
the edge $\{i ,k_1\}$,

$T_j^{1} \equiv$ subtree in $T_1$ rooted at $j$ except for the branch which contains
the edge $\{j ,k_m\}$,

$T_{k_r} \equiv$ subtree in $T_1$ rooted at $k_r$ except for the branches which
contain
either of the edges $\{k_{r-1} ,k_r\}$ and $\{k_r ,k_{r+1}\}$ ($k_0=i, k_{m+1}=j$),

$T_i^{2} \equiv$ subtree in $T_2$ rooted at $i$ except for the branch which contains
the edge $\{i ,j\}$, and

$T_j^{2} \equiv$ subtree in $T_2$ rooted at $j$ except for the branch which contains
the edge $\{i ,j\}$.

It is easy to see that the set $T_i^{2} \bigcup T_j^{2}$ is nonempty as
$T_1$ and $T_2$, being distinct, must contain more than two vertices.
Also $T_i^{2}$ and $ T_j^{2}$ must be disjoint; for, otherwise there will be
a path between $i$ and $j$ in $T_2$ which does not contain the edge $\{i,j\}$.
Thus there will be two paths between $i$ and $j$ in $T_2$
contradicting the fact
that $T_2$ is a spanning EPR tree (Figure \ref{figure37}).
With these two charactistics of $T_i^{2}$ and $ T_j^{2}$,
it is clear that $k_1$ will lie either in $T_i^{2}$ or in $ T_j^{2}$.
Without loss of generality, let us assume that $k_1 \in T_i^{2}$.
Now we do bicolored merging where the color $A$ is assigned to
$i$ and all vertices in $T_i^{1}$ and the color $B$ is assigned to
the rest of the vertices (refer to Figure \ref{figure37} for
illustration).
Since $T_1$ and $T_2$ were choosen arbitrarily,
the same arguments also imply that there can not exist
a method which converts $T_2$ to $T_1$.
This leads to the conclusion that any two distinct spanning EPR trees are
LOCC incomparable.

\begin{corollary}
\label{treenum}
There are at least exponentially many LOCC-incomparable classes of
pure multi-partite entangled states.
\end{corollary}
{\it Proof:}
We know from results in graph theory \cite{ndeo} that on a labelled 
graph on $n$ vertices, there are $n^{n-2}$ posible distinct spanning trees.
Hence there are $n^{n-2}$ distinct spanning EPR trees 
in a network of $n$ agents.
>From Theorem \ref{twoeprtrees} all these spanning 
EPR trees are LOCC incomparable.
It can be noted here that the most general local 
operation of $n$ qubits is an element of the group $U(2)^n$ 
(local unitary rotations on each qubit alone). 
So, if two states are found incomparable, 
this means that there are actually two incomparable equivalence 
classes of states (where members in a class are related by a $U(2)^n$ transformation).
Thus we have at least exponentially many LOCC-incomparable 
classes of multi-partite entangled states. \hfill \qed

\section{Combinatorial conditions for LOCC incomparability of entangled hypergraphs}

Since  entangled  hypergraphs   represent  more  general  entanglement
structures  than those represented  by the  EPR graphs  (in particular
spanning EPR trees are nothing but 2-uniform entangled hypertrees), it
is  likely  that there  will  be  even  more classes  of  incomparable
multi-partite  states  and this  motivates  us  to generalize  Theorem
\ref{twoeprtrees} for entangled  hypertrees.  However, remarkably this
intuition does  not work directly  and there are  entangled hypertrees
which are not incomparable.  But there are a large number of entangled
hypertrees which do not fall  under any such partial ordering and thus
remain incomparable. To this  end we present our first imcomparability
result on entangled hypergraphs.

\begin{theorem}
\label{pendhypincomp}
Let $H_1=(S,F_1)$ and $H_2=(S,F_2)$ be two entangled hypertrees.
Let $P_1$ and $P_2$ be the set of pendant vertices of $H_1$ 
and $H_2$ respectively. 
If the sets $P_1 \setminus P_2$ and $P_2 \setminus P_1$ are both 
nonempty then the multi-partite states represented by $H_1$ and $H_2$ are
necessarily LOCC-incomparable.
\end{theorem} 

\begin{figure}
\scalebox{0.4}{\includegraphics{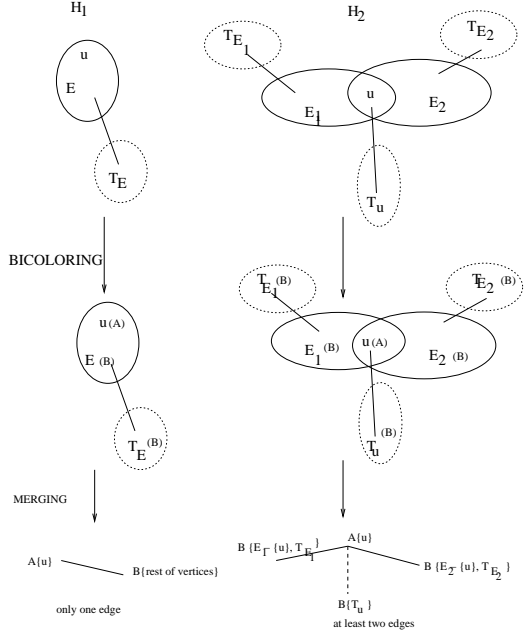}}
\caption{Entangled hypergraphs with $P_1 \setminus P_2$ non-empty}
\label{figure30D}
\end{figure}

Proof: Using bicolored merging we first show that
$H_1$ can not be converted to $H_2$ under LOCC.
Impossibility of the reverse conversion will also be immediate.
Since $P_1 \setminus P_2$ is nonempty,
there exists $u \in S$ such that $u \in P_1 \setminus P_2$.
That is, $u$ is pendant in $H_1$ but non-pendant
in $H_2$ (Figure \ref{figure30D}).

In the bicolored merging assign the color $A$
to the vertex $u$ and the color $B$ to all other vertices.
This reduces $H_1$ to a single EPR pair shared
between the two parties $A$ and $B$ whereas
$H_2$ reduces to at least two EPR pairs shared between $A$ and $B$.
The complete bicolored merging is
shown in Figure \ref{figure30D}. \hfill \qed

We note that this proof does not utilize the fact that
$H_1$ and $H_2$ are entangled hypertrees, and thus
the theorem is indeed true even for entangled hypergraphs
satisfying the conditions specified on the set of pendant vertices.

\begin{figure}
\scalebox{0.4}{\includegraphics{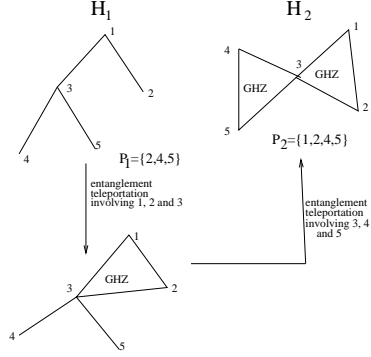}}
\caption{Comparable with $P_1 \neq P_2$ and $P_1 \subset P_2$}
\label{figure38}
\end{figure}

\begin{figure}
\scalebox{0.3}{\includegraphics{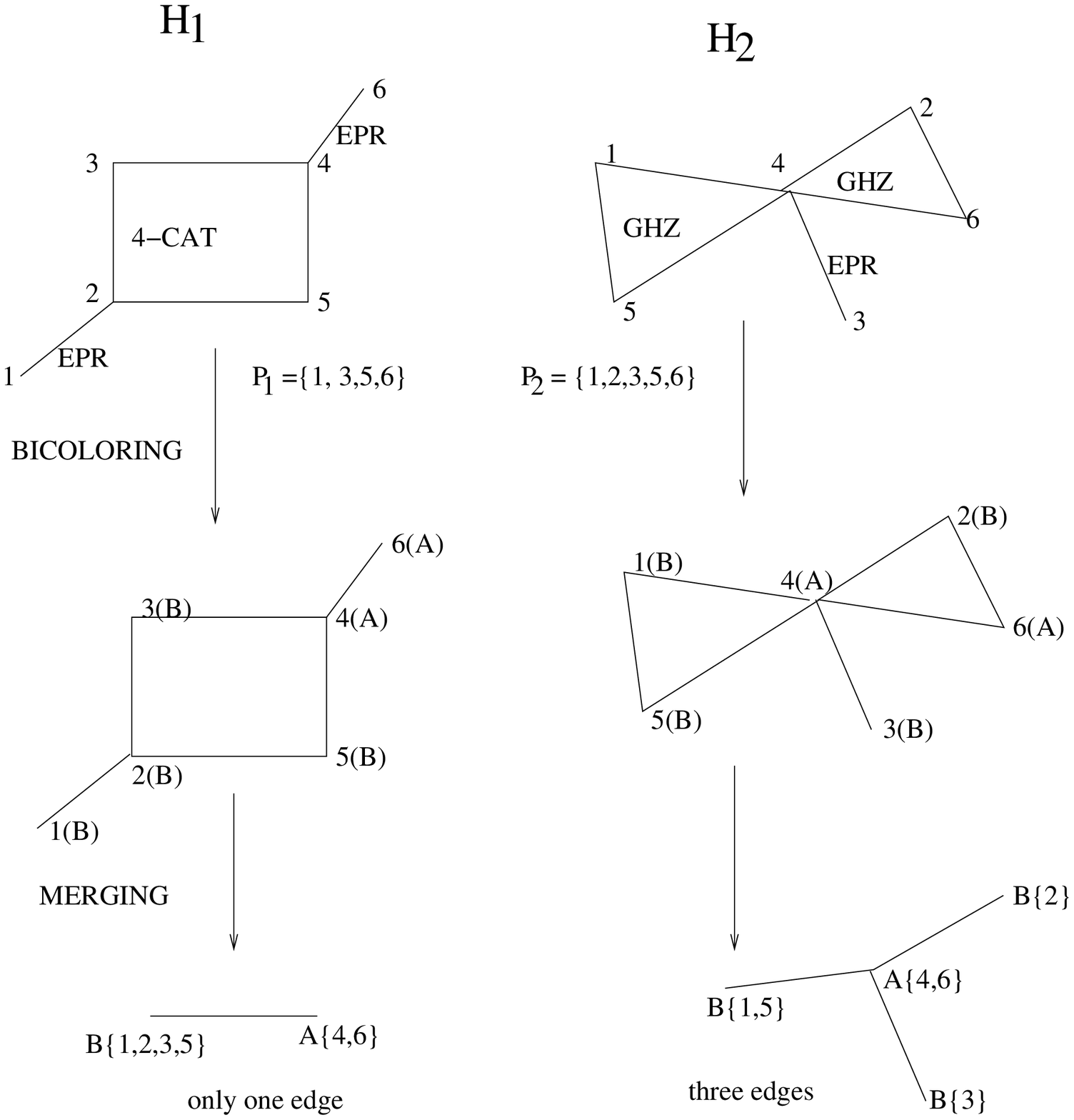}}
\caption{InComparable with $P_1 \neq P_2$ and $P_1 \subset P_2$}
\label{figure39}
\end{figure}

\begin{figure}[htbp]
\scalebox{0.4}{\includegraphics{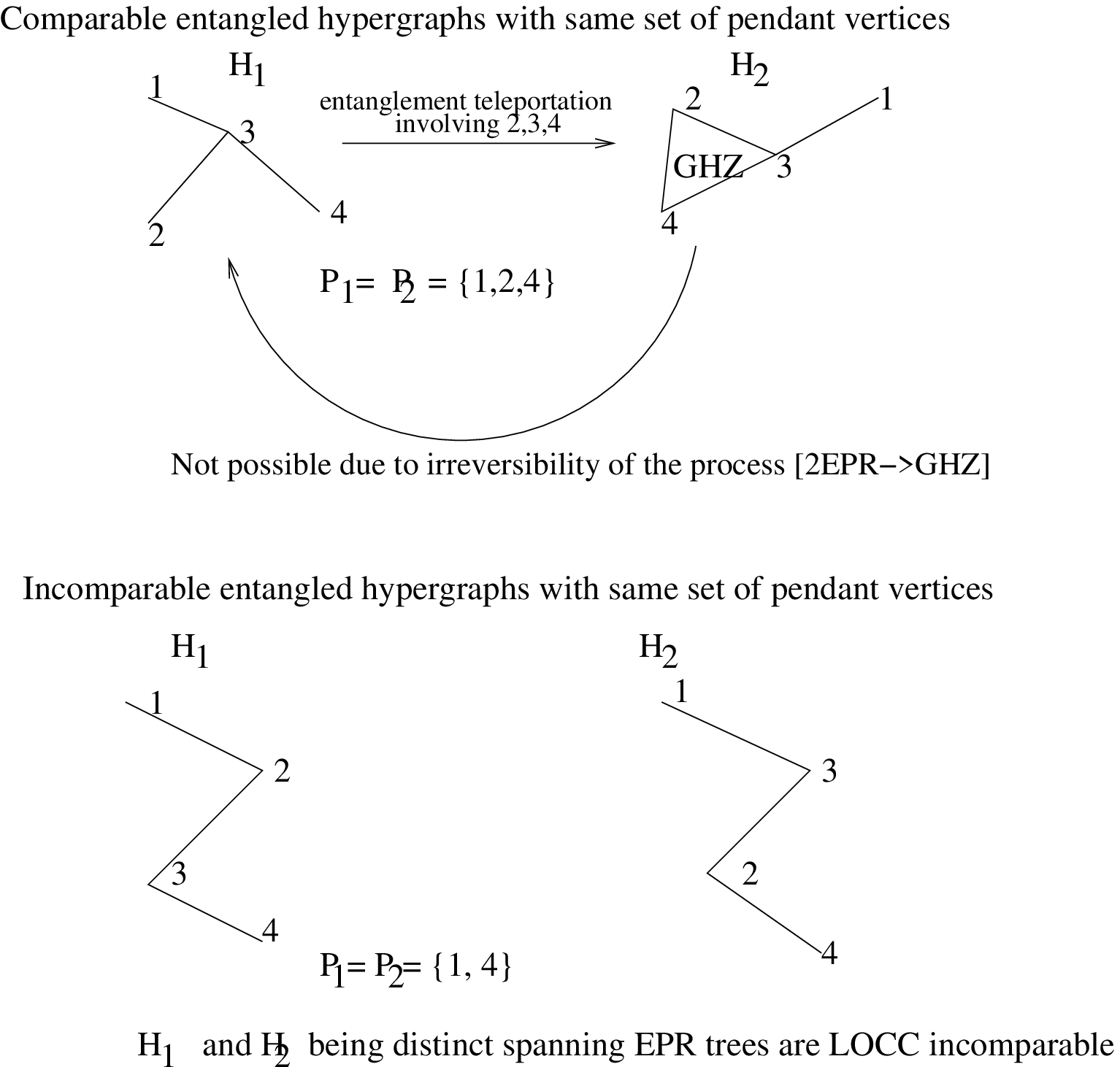}}
\caption{$P_1 = P_2$}
\label{figure310}
\end{figure}

\begin{figure}
\scalebox{0.4}{\includegraphics{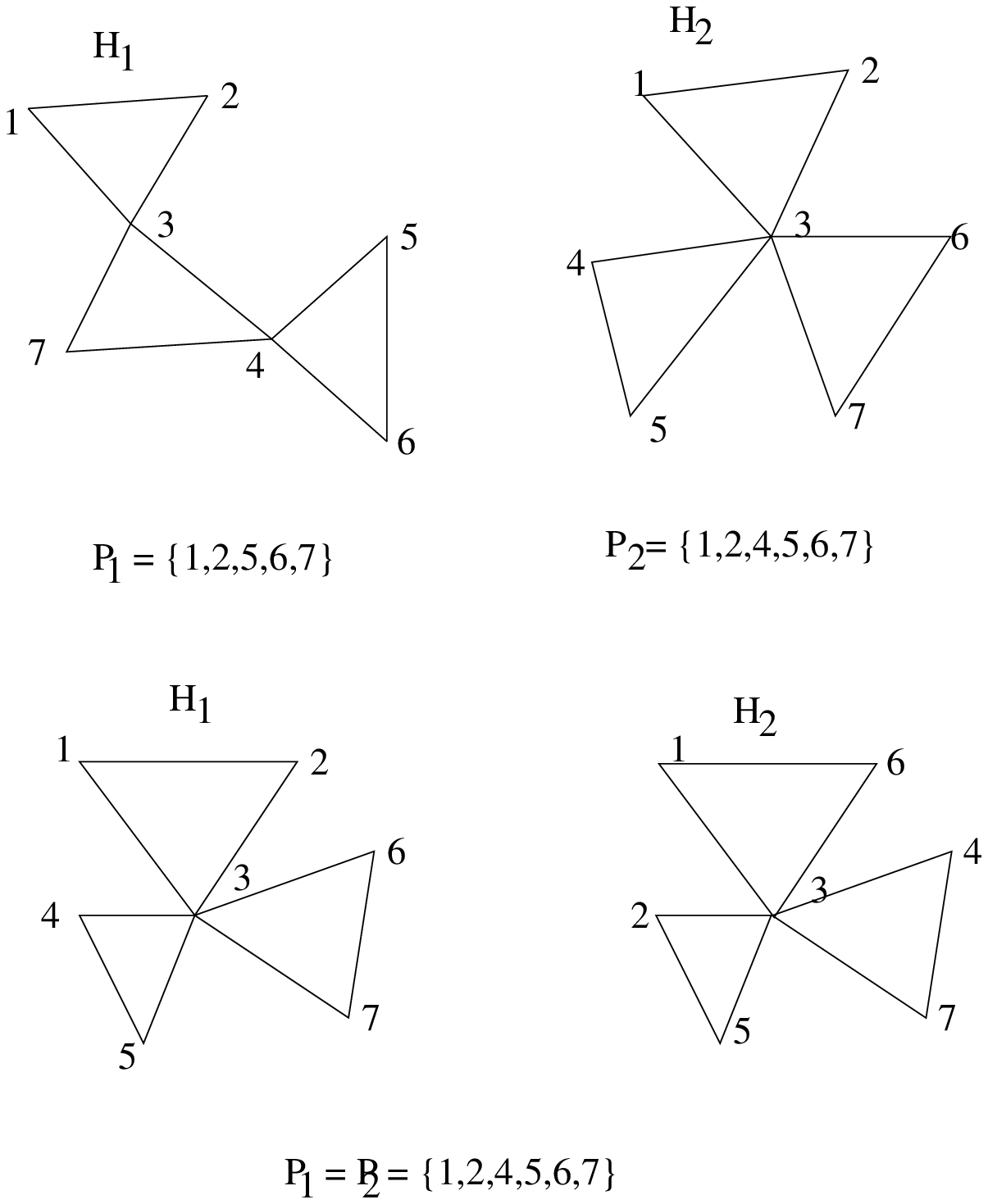}}
\caption{$r$-uniform entangled hypertrees not captured in Theorem
\ref{pendhypincomp}}
\label{figure311}
\end{figure}

The conditions specified on the set of pendant vertices in Theorem
\ref{pendhypincomp} cover a very small fraction of the entangled hypergraphs.
However, these conditions are not necessary and it may be possible to 
find further characterizations of incomparable classes 
of entangled hypergraphs. We present two examples where the conditions 
of Theorem \ref{pendhypincomp} are not satisfied. 

{\it Example-1}:(Figures \ref{figure38} and \ref{figure39})
$P_1 \neq P_2$ but either $P_1 \subset P_2$ or $P_2 \subset P_1$.

{\it Example-2}: (Figure \ref{figure310}) $P_1 = P_2$

In the first example, 
the entangled hypergraphs $H_1$ and $H_2$ staisfy 
$P_1 \neq P_2$ and $P_1 \subset P_2$. 
$H_1$ and $H_2$ are comparable in Figure \ref{figure38} 
but incomparable in Figure \ref{figure39}.
In Figure \ref{figure39}, the incomparability has been proved by
showing that $H_1$ is not convertible to $H_2$ under LOCC because
the impossibility of reverse conversion follows from the proof of 
Theorem \ref{pendhypincomp} ($P_2 \setminus P_1 \neq \phi$).
Figure \ref{figure310} gives examples of comparable and 
incomparable entangled hypergraphs
with condition $P_1 = P_2$.  
 
Theorem \ref{twoeprtrees}  shows that two distinct  EPR spanning trees
are  LOCC incomparable  and the  spanning  EPR trees  are nothing  but
$2$-uniform entangled hypertrees.  Therefore, a natural generalization
of this theorem  would be to $r$-uniform entangled  hypertrees for any
$r \geq  3$.  As we show  below, the generalization  indeed holds.  It
should be noted that  Theorem \ref{pendhypincomp} does not necessarily
capture  such entanglement  structures (multi-partite  states) (Figure
\ref{figure311}).   However,  in  order  to prove  that  two  distinct
$r$-uniform entangled  hypertrees are  LOCC incomparable, we  need the
following important result  about $r$-uniform hypertrees. See Appendix
2 for the proof.
\begin{theorem}
\label{lemmaruht}
Given   two   distinct   $r$-uniform  hypertrees   $H_1=(S,F_1)$   and
$H_2=(S,F_2)$ with $r \geq 3$, there exist vertices $ u, v \in S$ such
that $u$ and $v$ belong to  same hyperedge in $H_2$ but necessarily to
different hyperedges in $H_1$.
\end{theorem}

Now  we state  one  of our  main  results on  LOCC incomparability  of
multi-partite entangled states in the following theorem.

\begin{theorem}
\label{hyptree}
Any two distinct $r$-uniform entangled hypertrees are LOCC-incomparable.
\end{theorem}
  
Proof: Let $H_1=(S,F_1)$ and $H_2=(S,F_2)$ be the two $r$-uniform entangled hypertrees.
If $r=2$ then $H_1$ and $H_2$ happen to be two distinct 
spanning EPR trees and the proof follows from the theorem \ref{twoeprtrees}.
Therefore, let $r \geq 3$.

Now from Theorem \ref{lemmaruht}, there exist $u,v \in S$ such that 
$u$ and $v$ belong to the same hyperedge in $H_2$
but necessarily to different hyperedges in $H_1$.
Let the same hyperedge in $H_2$ be $E \in F_2$. 
Also, since $H_1$, being hypertree, is {\it connected}, 
there exists a path between $u$ and $v$ in $H_1$.
Let this path be $u E_1 E_2 \cdots E_{k+1} v$.
Clearly $ k > 0$ because $u$ and $v$ necessarily do not belong 
to the same hyperedge in $H_1$. 

\begin{figure}
\scalebox{0.4}{\includegraphics{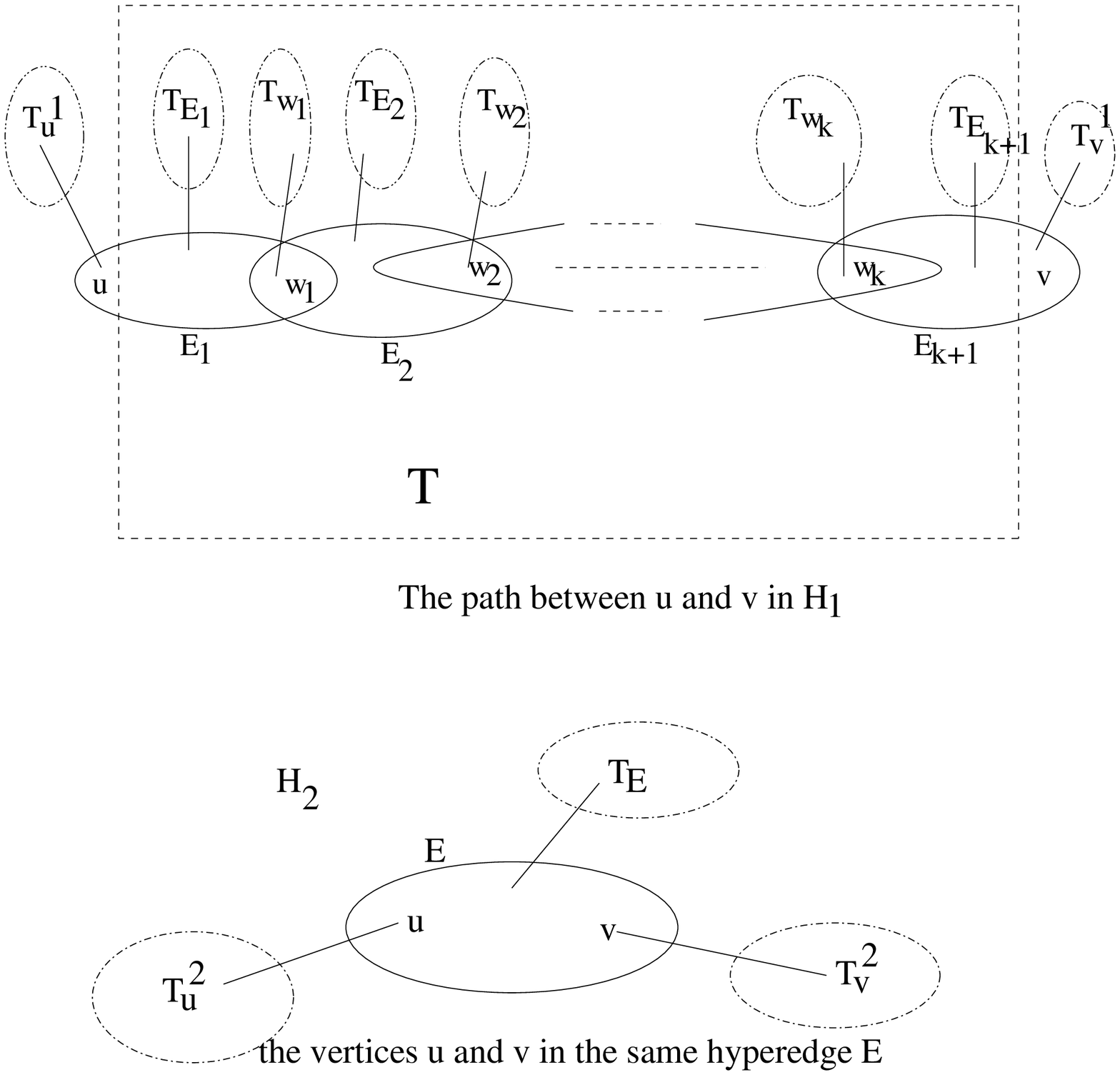}}
\caption{Two distinct $r$-uniform entangled hypertrees}
\label{figure312}
\end{figure}

We introduce the following notations (Figure \ref{figure312}).  

$T_u^{1} :$ sub-hypertree rooted at $u$ in $H_1$ except the branch 
that contains $E_1$.

$T_v^{1} :$ sub-hypertree rooted at $v$ in $H_1$ except the branch 
that contains $E_{k+1}$. 

$T_{w_i}:$ sub-hypertree rooted at $w_i$ in $H_1$ except 
branches containing
           $E_i$ and $E_{i+1}$.

$T_{E_i}:$ Collection of all sub-hypertrees in $H_1$ rooted at some vertices in $E_i$ 
        other than $w_{i-1}$ and $w_i$ (where $w_0 = u$ and $w_{k+1} = v$) 
        except for the branches which contain $E_i$.

$T = (( E_1 \bigcup E_2 \bigcup \cdots \bigcup E_{k+1}) \bigcup 
     ( T_{E_1} \bigcup T_{E_2} \bigcup \cdots \bigcup T_{E_{k+1}}) \bigcup
     ( T_{w_1} \bigcup T_{w_2} \bigcup \cdots \bigcup T_{w_k})) \setminus \{u,v\}$

   $=$ set of all vertices from $S \setminus \{u,v\}$ which are not contained in
   $T_u \bigcup T_v$. 

$T_u^{2} :$ sub-hypertree rooted at $u$ in $H_2$ except the
branch that contains $E$.

$T_v^{2} :$ sub-hypertree rooted at $v$ in $H_2$ except the branch that
contains $E$.

$T_E:$ Collection of all sub-hypertrees in $H_2$ rooted at some vertices in
       $E \setminus \{u,v\} $ except for the branches which contain $E$. 

In order to complete the proof we consider the following cases:

{\it CASE $1$}: $\exists w \in T$ such that 
$w \in (T_u^{2} \bigcup T_v^{2})$ 

Without loss of generality let us take $w \in T_u^{2}$.
Now since $w \in T$, $ w \in$ exactly one of $E_i$, 
$T_{w_i}$, or $T_{E_i}$ for some $i$. 
Accordingly there will be three subcases.
{\it Case $1.1$}: $w \in E_i$ for some $i$ (take such minimum $i$).

Do bicolored merging where the vertex $u$ along with all the vertices in 
\bigskip

$T_u^{1}, E_1, E_2, \cdots , E_{i-1}, T_{w_1}, T_{w_2}, \cdots , T_{w_{i-1}}, 
T_{E_1}, T_{E_2}, \cdots , T_{E_{i-1}}$ 

\bigskip

are given the color $A$ and the rest of the vertices are
given the color $B$.

{\it CASE $1.2$ }: $w \in T_{w_i}$ for some $i$. 

Do the bicolored merging while assigning the colors as in the above case.

{\it CASE $1.3$}: $w \in T_{E_i}$ for some $i$.

Bicolored merging in this case is also same as in CASE $1.1$.

{\it CASE $2$}: There does not exist any $w \in T$ 
such that $w \in T_u^{2} \bigcup T_v^{2}$.

Clearly, $ T_u^{2} \bigcup T_v^{2} \subset T_u^{1} \bigcup T_v^{1}$ and $T \subset T_E \bigcup (E \setminus \{u, v\})$.
Note that whenever we are talking of set relations like union, containment etc., we are considering 
the trees, edges etc. as sets of appropriate vertices from $S$ which make them.
First we establish the following claim.

{\it Claim}: $ \exists t \in (E_1 \setminus \{u, w_1\}) \bigcup (E_2 \setminus \{w_1, w_2\})$ such that
$ t \in T_E$. 

We have $ k > 0$.
Therefore, both $E_1$ and $E_2$ exist and since $H_1$ is $r$-uniform 
$|E_1| = |E_2| = r$. 
Also $ (E_1 \setminus \{u, w_1\}) \bigcap (E_2 \setminus \{w_1, w_2\}) $ 
is empty, for, otherwise 
there will be a cycle in $H_1$ which is not possible as $H_1$ is a hypertree \cite{hbc1,berge}.      
Therefore, 

\bigskip

$|(E_1 \setminus \{u, w_1\} \bigcup (E_2 \setminus \{w_1, w_2\})|
           = |(E_1 \setminus \{u, w_1\}| + |E_2 \setminus \{w_1, w_2\}|
           = (r-2) + (r-2) = 2r-4$. 
\bigskip

Also $|E| =r $ implies that $|E \setminus \{u,v\}| = (r-2)$.

It is clear that $ u, v \notin (E_1 \setminus \{u,w_1\}) \bigcup (E_2 \setminus \{w_1, w_2\})$.
\bigskip
Therefore,

$|(E_1 \setminus \{u,w_1\}) \bigcup (E_2 \setminus \{w_1, w_2\})| - |E \setminus \{u,v\}|
=(2r-4)-(r-2) = r-2 \geq 1 $ since $ r \geq 3$.
\bigskip

Also $(E_1 \setminus \{u,w_1\}) \bigcup (E_2 \setminus \{w_1 , w_2\}) 
\subset T \subset T_E \bigcup (E \setminus \{u,v\})$,
\bigskip

and so by Pigeonhole principle \cite{pigeon},
\bigskip

 $ \exists t \in (E_1 \setminus \{u,w_1\}) \bigcup (E_2 \setminus \{w_1,w_2\})$
and
$t \in T_E (\notin (E \setminus \{u,v\}))$.

Hence our claim is true.   

Now we have $t \in (E_1 \setminus \{u,w_1\}) \bigcup (E_2 \setminus \{w_1,w_2\})$ such that $t \in T_E$.
Since $t \in T_E$, by the definition of $T_E$ it is clear that there must exist $w \in E \setminus \{u,v\}$
such that $ t \in T_w$, the sub-hypertree in $H_2$ rooted at $w$ except for the branch containing $E$.
Depending on whether $t \in E_1 \setminus \{u,w_1\}$ or $t \in  E_2 \setminus \{w_1,w_2\}$,
we break this case into several subcases and futher in sub-subcases depending on the part in $H_1$
where $w$ lies. 

\begin{figure}
\scalebox{0.4}{\includegraphics{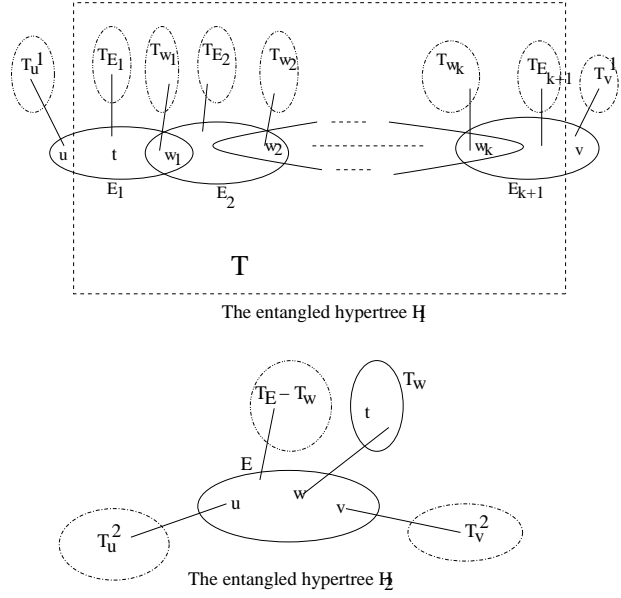}}
\caption{CASE $2.1$}
\label{figure313}
\end{figure}

{\it CASE $2.1$}: $ t \in E_1 \setminus \{u,w_1\}$ (Figure \ref{figure313}). 

{\it CASE $2.1.1$}: $ w \in T_u^{1}$.

Do the bicolored merging where $u$ and the vertices in 
$T_u^{1}$ are assigned the color $A$ and 
the rest of the vertices from $S$ are given the color $B$.     
   
{\it CASE $2.1.2$}: $ w \in T_v^{1} $.

Bicolored merging is done where $v$ as well as all the vertices in $T_v^{1}$ are assigned the color 
$B$ and rest of the vertices from $S$ are given the color $A$. 

{\it CASE $2.1.3$ }: $w \in T $.

Here in this case, depending on whether $w$ is in $T_t$ or 
not, there can be two cases.

{\it CASE $2.1.3.1$}: $ w \in T_t$.

Bicolored merging is done where all the vertices in $T_t$ are given the color $A$ and
rest of the vertices are assigned the color $B$.    

{\it CASE $2.1.3.2$}: $ w \notin T_t$.

$w \notin T_t$ implies that either $w \in E_i$ for some $i$,
or $ w \in T_q$, where $ q \in E_i$ 
for some $i$ and $ q \neq t$. 
For both of these possibilities, bicolored 
merging is the same and is done as follows:

Assign the color $A$ to $u$ as well as all vertices in
 
$T_u^{1} \bigcup E_1 \bigcup T_{E_1} \bigcup  T_{w_1} \bigcup \cdots \bigcup E_{i-1} \bigcup T_{E_{i-1}} 
\bigcup T_{w_{i-1}} \bigcup (E_i \setminus \{q,w,w_i\}) \bigcup (T_{E_i} \setminus T_q)$ 

and assign the color $B$ to rest of the vertices. 

\begin{figure}
\scalebox{0.5}{\includegraphics{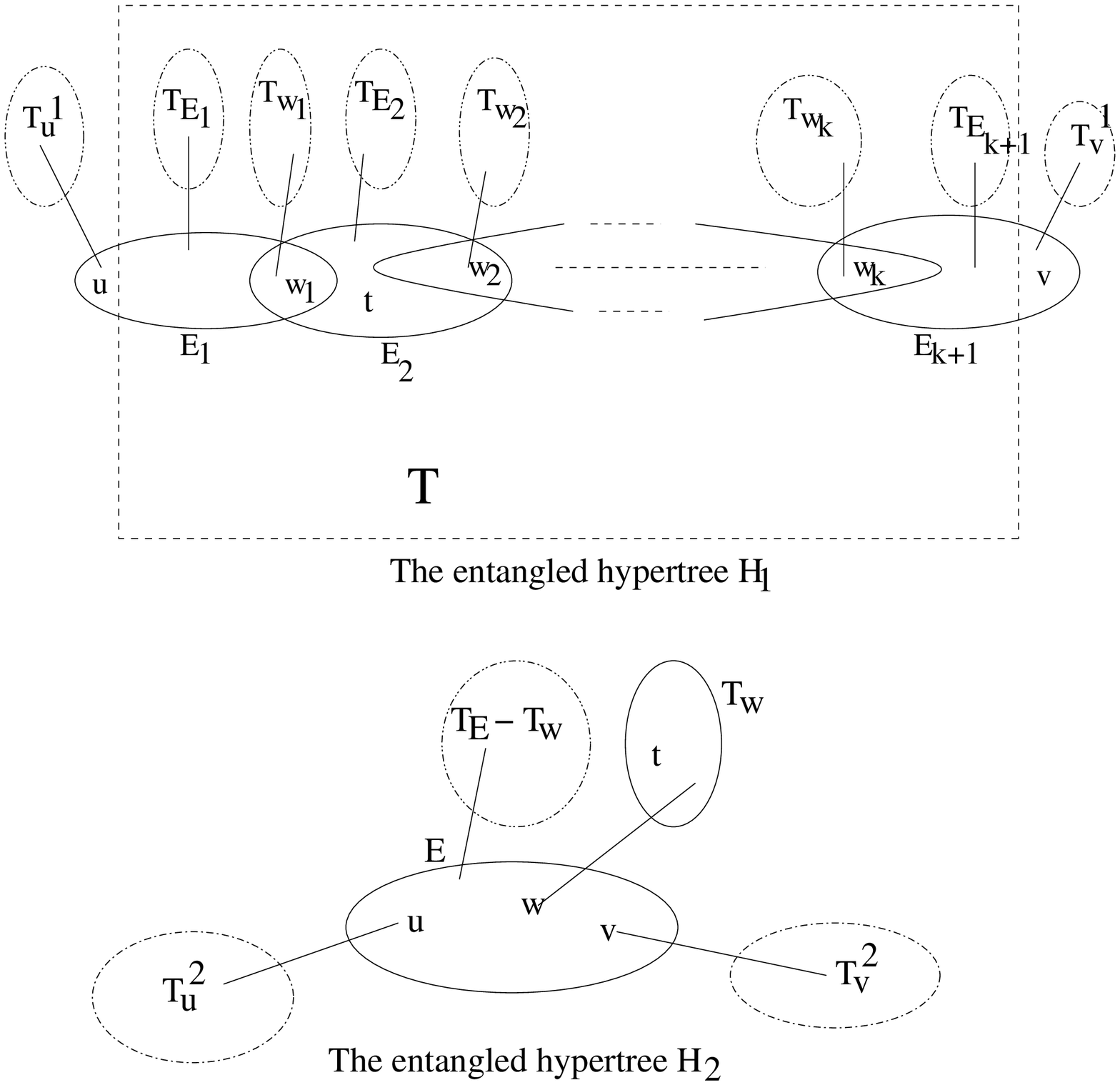}}
\caption{CASE $2.2$}
\label{figure314}
\end{figure}

{\it CASE $2.2$}: $t \in E_2 \setminus \{w_1,w_2\}$ (Figure \ref{figure314}).

{\it CASE $2.2.1$}: $w \in T_u^{1} \bigcup E_1 \bigcup T_{E_1} \bigcup T_{w_1}$.

Do the bicolored merging where all the vertices in $T_u^{1} \bigcup E_1 \bigcup T_{E_1} \bigcup T_{w_1}$
including $u$  
are given the color $A$ and rest of the vertices are assigned the color $B$. 

{\it CASE $2.2.2$}: $ w \in T_v^{1} \bigcup T_{E_{k+1}} \bigcup E_{k+1} \bigcup T_{w_k} \bigcup \cdots \bigcup T_{E_3} \bigcup E_3 \bigcup T_{w_2}$.

In bicolored merging give the color $B$ to all the vertices (including $v$) in 

$ T_v^{1} \bigcup T_{E_{k+1}} \bigcup E_{k+1} \bigcup T_{w_k} \bigcup \cdots \bigcup T_{E_3} \bigcup E_3 \bigcup T_{w_2}$
 
and color $A$ to the rest of the vertices.  

{\it CASE $2.2.3$}: $ w \in E_2 \bigcup T_{E_2}$. 

In this case depending on whether $w \in T_t$, or $w \notin T_t$ the bicolored merging will be different.

{ \it CASE $2.2.3.1$}: $w \in T_t$.

Bicolored merging is done where all the vertices in $T_t$ are given the color $A$ and 
rest of the vertices are assigned the color $B$.

{\it CASE $2.2.3.2$}: $ w \notin T_t$.

$ w \notin T_t $ implies that either $w \in E_2$,
or $ w \in T_q $ for some $q ( \neq t)  \in E_2$.
In any case do the bicolored merging where the color $A$ is assigned to all the vertices in

$ T_u^{1} \bigcup E_1 \bigcup T_{E_1} \bigcup T_{w_1} \bigcup (E_2 \setminus \{w,q,w_2\})
\bigcup (T_{E_2} \setminus T_q)$ 

and rest of the vertices are assigned the color $B$.

Now that we have exhausted all possible cases and shown by the method of bicolored merging that
the $r$-uniform entangled hypertree $ H_1$ can not be LOCC converted to the $r$-uniform entangled 
hypertree $H_2$. The same arguments also work for showing that $H_2$ can not be LOCC converted to
$H_1$ by interchanging the roles of $H_1$ and $H_2$. 
Hence the theorem follows. \hfill \qed 

\begin{figure}
\scalebox{0.4}{\includegraphics{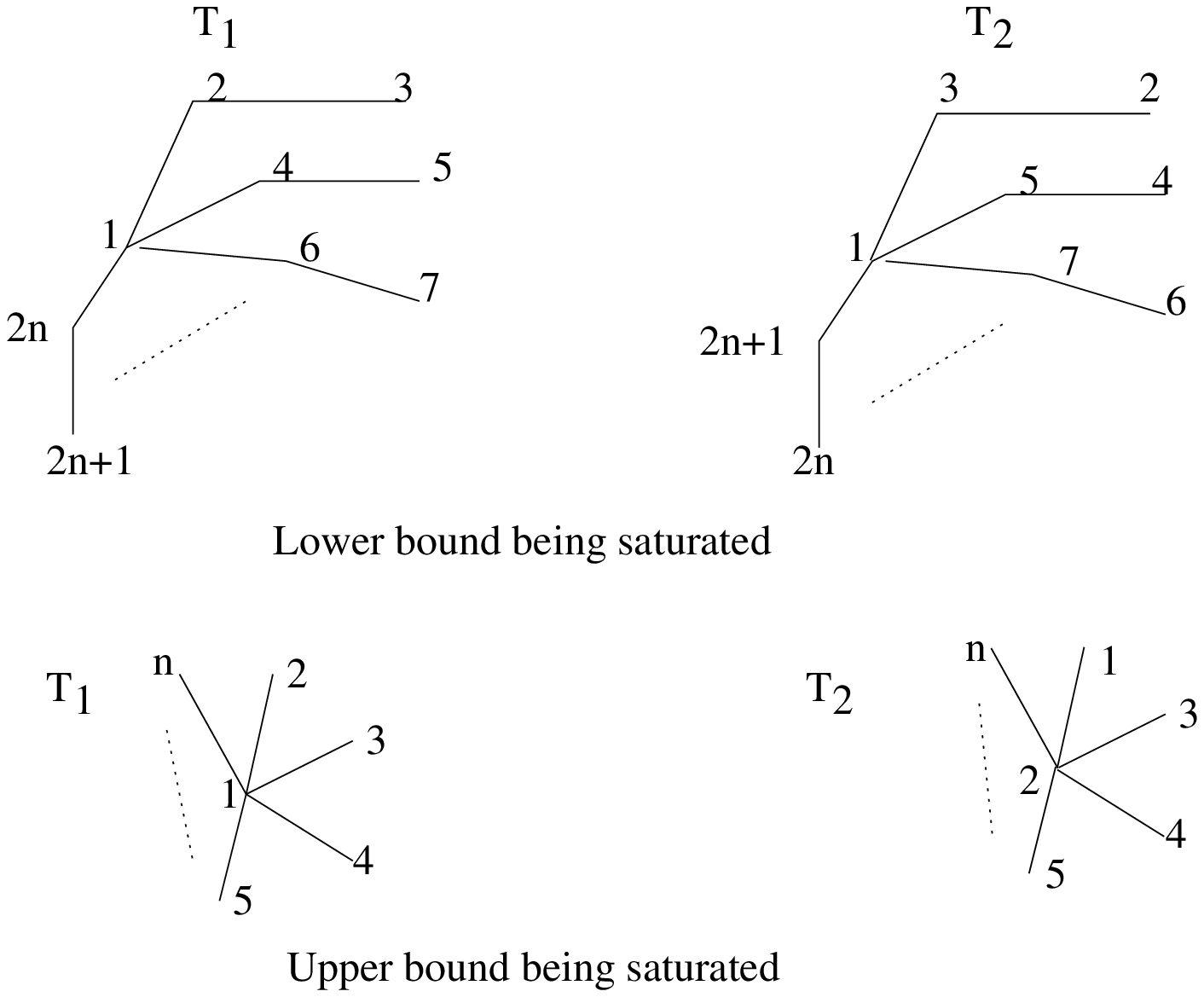}}
\caption{}
\label{figure315}
\end{figure}

Before ending our section on LOCC incomparability of multi-partite states 
represented by EPR graphs and entangled hypergraphs, we note that 
 partial entropic criteria of Bennett et. al\cite{bennet2000} 
which gives a sufficient condition for 
LOCC incomparability of multi-partite states, does not capture
the LOCC-incomparability of spanning EPR trees 
or entangled hypertrees in general.
Consider two spanning EPR trees $T_1$ and $T_2$
on three vertices (say $1, 2, 3$). 
$T_1$ is such that the vertex pairs $ 1, 2$ and $1,3$ are forming 
the two edges where as in $T_2$ the vertex pairs $1,3$ and $2,3$
are forming the two edges.  
It is easy to see that $T_1$ and $T_2$ are not marginally isentropic.  

\section{Quantum Distance between Multi-partite Entangled States} 
\label{quantumdistance}

In the proof of Theorem \ref{twoeprtrees}, we have utilized the fact that
there exist at least two vertices which are connected by an edge in $T_2$ 
but not in $T_1$.
This follows as $T_1$ and $T_2$ are different and
they also have equal number of edges (namely $n-1$, if there are $n$ vertices).
In fact, in general there may exist several such pairs of vertices depending on
the structures of $T_1$ and $T_2$. Fortunately, 
the number of such pair of vertices has some nice 
features giving rise to a metric on the set 
of spanning (EPR) trees with fixed vertex set 
and thus giving a concept of distance 
\cite{ndeo}.
The distance between any two spanning (EPR) trees $T_1$ and $T_2$ denoted
by $QD_{T_1,T_2}$ on the same vertex set is defined as 
the number of edges in $T_1$ which are not in $T_2$.         
Let us call this distance to be the {\it quantum distance} between $T_1$ and $T_2$.
We have proved in Theorem \ref{twoeprtrees} that obtaining $T_2$ from $T_1$ 
is not possible just through LOCC, so we need to do quantum communication.
The minimum number of qubit required to be communicated for 
this purpose should be an interesting parameter related to state transformations
amongst multi-partite states represented by spanning EPR trees; let us
denote this number by $q_{T_1,T_2}$.
We note that $ q_{T_1,T_2} \leq QD_{T_1,T_2}$. This is because 
each edge not present in $T_2$ can be created by only one qubit communication.
The exact value of $q_{T_1,T_2}$ will depend on the structures of $T_1$ and $T_2$ 
and, as we can note, on the number of edge disjoint paths in $T_1$ between the 
vertex pairs which form an edge in $T_2$ but not $T_1$.

We can say more about quantum distance.
Recall Theorem \ref{treecat} where we show that 
a lower bound on the number of copies of $n$-CAT  
to prepare a spanning EPR tree by LOCC, is $n-1$. 
Can we obtain a similar lower bound in the case of two 
spanning EPR trees and relate it to the quantum distance?
The answer is indeed yes. 
Let $C_{T_1,T_2}$ denote the minimum number of copies of
the spanning EPR tree $T_1$ required to obtain $T_2$ just by LOCC. 
We claim that $ 2 \leq C_{T_1,T_2}$ , $C_{T_2,T_1} \leq QD_{T_1,T_2} + 1 $.
The lower bound follows from Theorem \ref{twoeprtrees}.
The upperbound is also true because of the following reason.
$QD_{T_1,T_2}$ is the number of (EPR pairs) edges present in $T_2$ but not in 
$T_1$. For each such edge in $T_2$ (let $u, v$ be the vertices forming the edge),
while converting many copies of $T_1$ to $T_2$ by LOCC
an edge between $u$ and $v$ must be created. 
Since $T_1$ is a spanning tree and therefore connected, there must be a path
between $u$ and $v$ in $T_1$ and this path can be well converted (using entanglement 
swapping) to an edge between 
them ( i.e. EPR pair between them) only using LOCC. Hence one copy each will suffice to
create each such edges in $T_2$. Thus $QD_{T_1,T_2}$ copies of $T_1$ will be sufficient 
to create all such $QD_{T_1,T_2}$ edges in $T_2$. One more copy will supply all the edges 
common in $T_1$ and $T_2$.  
Even more interesting point is that both these bounds are saturated. This means to say that
there do exist spanning EPR trees satifying these bounds (Figure \ref{figure315}).

It is important to note that a similar concept of distance also holds in the case of
$r$-uniform entangled hypertrees.
  
\begin{acknowledgments}
We thank P. Panigrahi, S. P. Khastgir and K. Mitra for discussions. 
SKS thanks Prof. V. P. Roychoudhury for providing an important reference.
\end{acknowledgments}

~\\
\hrule
~\\

\newpage

\appendix{\bf Appendix 1}

\noindent
{\bf Proof of Theorem \ref{skpnec}}: \\ We use the method of bicolored
merging  to prove  the fact  that any  disconnected EPR  graph  $G$ on
$n$-vertices  can  not be  converted  to  an  $n$-CAT state  on  those
vertices  under LOCC.  We first  note  that the  BCM EPR  graph of  an
$n$-CAT state, irrespective of the  bicoloring done, is always a graph
which contains  exactly one edge. Now  as $G$ is  disconnected it will
have more than one connected components. Let these components be $C_1,
C_2 \vdots C_k$, where $k \geq 2$ . The bicoloring is done as follows:
assign the  color $A$ to all  the vertices in the  component $C_1$ and
the color $B$  to all other vertices i.e.  all vertices in $G\setminus
C_1$. After  merging, therefore, $G$  reduces to a  disconnected graph
with  no edges  i.e. the  BCM EPR  graph of  $G$ is  a graph  with $k$
isolated  vertices and  no edges.  Now if  we are  able to  prepare an
$n$-CAT state from  $G$ just using LOCC, we could  also prepare an EPR
pair between two parties who were never sharing an EPR pair just using
LOCC.   This  violates   monotonicity   and  hence   the  theorem   is
proved. \hfill \qed
\bigskip

\appendix{\bf Appendix 2} \label{apx2}

\noindent
{\bf Proof of Theorem \ref{lemmaruht}}: \\
We first establish the following claim: 

{\it Claim}: $ \exists E_1 \in F_1 , E_2 \in F_2$ such that $E_1 \bigcap E_2 
\neq \phi$ and $ E_2 \notin F_1 \bigcap F_2$.

{Proof of the claim:} We first show that on the same vertex set, the 
number of hyperedges in any $r$-uniform hypertree is always same. Let 
$n$ and $m$ be the number of vertices and hyperedges in a $r$-uniform 
hypertree. We show by induction on $m$ that $n = m*(r-1) + 1$.

For $m = 1$, $ n = 1 *(r-1) + 1 = r$ which is true because all possible 
vertices (since no one can be isolated) fall in the single hyperedge and 
it has exactly $r$ vertices.

Let us assume that this relation between $n$ and $m$ for a fixed $r$ 
holds for all values of the induction variable up to $m-1$. We show 
that it holds good for $m$.

Now take  a $r$-uniform hypertree  with $m$ hyperedges. Remove  any of
the hyperedges to get another  hypergraph (which may not be connected)
having  only $m-1$  edges. This  removal may  introduce  $k$ connected
components (sub-hypertrees); $ 1 \leq  k \leq r$. Let these components
have   respectively    $m_1,   m_2,   \cdots   ,    m_k$   number   of
hyperedges. Therefore,  $ \sum_{i=1}^{k} m_i = m-1$.  The total number
of  vertices in  the new  hypergraph (with  the $k$  sub-hypertrees as
components), $n^{1} = \sum n_i $ where $n_i$ is the number of vertices
in the  component $i$. Therefore,  $n^{1} = \sum n_i  = \sum_{i=1}^{k}
\{m_i (r-1) +1\} = (m-1)(r-1) + k$ (under induction assumption).

Now the number of vertices in  the original hypertree, $n = n^{1} + (r
- k)$ because $k$  vertices were already covered, one  each in the $k$
components. Therefore, $ n = (m-1)(r-1) + k + (r-k) = (m-1)(r-1) + r =
(m-1)(r-1) + (r-1) +1  = m(r-1) + 1$. The result is  thus true for $m$
and  hence for  any number  of  hyperedges by  induction. This  result
implies that  any $r$-uniform  hypertree on the  same vertex  set will
always have the same number of hyperedges.

Let $ F = F_1 \bigcap F_2$ and $  m = |F_1| = |F_2| $. Obviously $ m >
|F| $  otherwise $H_1 = H_2$. This  implies that $\exists E  \in F_2 $
such that $E \notin F$.

Take  any vertex  say  $w \in  E$.  Since $w  \in S$  and  $H_1$ is  a
hypertree therefore connected,  $w$ can not be an  isolated vertex and
therefore $\exists E^1 \in F_1$  such that $w \in E^1$. Take $E_1=E^1$
and $E_2=E$. This proves our claim.

Now we prove the theorem. Choose $E_1$ and $E_2$ so as to 
satify the above claim.

Let $E_1 = \{ u_1, u_2, \cdots , u_l, w_{l+1}, w_{l+2}, \cdots , w_r \}$ and

$E_2 = \{u_1, u_2, \cdots , u_l, v_{l+1}, v_{l+2}, \cdots , v_r\}$.

Since $ E_1 \bigcap E_2 \neq \phi $, $ l \geq 1$ and $E_1 \neq E_2$ 
implies that $ l \leq r-1$.

Hence $ 1 \leq l \leq r$.

Now based on the value of $l$, we have following different cases:

{\it CASE $1$ }: $ l > 1$

{\it CASE $1.1$}: $\exists v_i$ such that $u_1$ and $v_i$ are not in

same hyperedge in $H_1$.

Take $u=u_1$ and $v = v_i$ in the statement of the theorem.

{\it CASE $1.2$}: Each $v_i$ is in some hyperedge in

$H_1$ in which $u_1$ also lies.

None of these $v_i$'s can belong to the hyperedges in $H_1$ in which $u_2$
lies.

This is due to the fact that if, say, $v_j$ happens to be in same hyperedge
as of $u_2$ in $H_1$ then
$u_1 u_2 v_j u_1$ will be a cycle in $H_1$,
which is absurd as $H_1$ is a hypertree.

Note that at least one such $v_i$ must exist as $l < r$.
Take $u=u_2$ and $v =$ any $v_i$.

{\it CASE $2$}: $ l = 1$

{\it CASE $2.1$}: $ \exists v_i$ such that $u_1$ and $v_i$ are 
not in same hyperedge in $H_1$.

Take $ u = u_1$ and $ v = v_i$.

{\it CASE $2.2$}: Each $v_i$ is in some hyperedge in
$H_1$ in which $u_1$ also lies.

Since $v_i$'s are $r-1$ in number and $E_2 \notin F_1 \bigcap F_2$,
these $v_i$'s will be distributed in
at least two distinct hyperedges in $H_1$ in which $u_1$ also lies.

Therefore, $ \exists v_i, v_j$ such that they are in the same hyperedge
in $H_2$ (namely in $E_2$) but
in necessarily different edges in $H_1$, otherwise
(that is, if they lie in the same hyperedge in $H_1$)

$ u_1 v_i v_j u_1 $ will be a cycle in $H_1$,
which is absurd as $H_1$ is a hypertree.

Also note that both $v_i$ and $v_j$ will exist as $r \geq 3$.

Take $u = v_i$ and $v=v_j$.

Thus we have proved Theorem \ref{lemmaruht} in all possible cases. 

\hfill \qed

We would like to point out that the result
of Theorem \ref{lemmaruht} could follow from some
standard results in combinatorics.
We have however not found literature proving this result. 
\end{document}